\documentclass[twocolumn]{aastex63}
\usepackage{epsfig,graphicx,rotating}
\def\lap{\lower.5ex\hbox{$\; \buildrel < \over \sim \;$}}
\def\gap{\lower.5ex\hbox{$\; \buildrel > \over \sim \;$}}

\def\ergcm2s{${\rm erg\ cm^{-2}\ s^{-1}}$}

\def\ergscm2s{${\rm erg\ cm^{-2}\  s^{-1}}$}

\def\cm-2{${\rm cm^{-2}}$}

\defcitealias{williams2014}{W14}

\begin{document}

\title{The Panchromatic Hubble Andromeda Treasury  XXI. The Legacy Resolved Stellar Photometry Catalog}

\author[0000-0002-7502-0597]{Benjamin F. Williams}
\affiliation{Department of Astronomy, University of Washington, Box 351580, U.W., Seattle, WA 98195-1580, USA}

\author[0000-0001-7531-9815]{Meredith Durbin}
\affiliation{Department of Astronomy, University of Washington, Box 351580, U.W., Seattle, WA 98195-1580, USA}

\author[0000-0002-1172-0754]{Dustin Lang}
\affiliation{McWilliams Center for Cosmology, Department of Physics, Carnegie Mellon University, 5000 Forbes Ave., Pittsburgh, PA}

\author[0000-0002-1264-2006]{Julianne J. Dalcanton}
\affiliation{Department of Astronomy, University of Washington, Box 351580, U.W., Seattle, WA 98195-1580, USA}
\affiliation{Center for Computational Astrophysics, Flatiron Institute, 162 Fifth Ave, New York, NY 10010, USA}

\author[0000-0001-8416-4093]{Andrew E. Dolphin}
\affiliation{Raytheon, 1151 E. Hermans Road, Tucson, AZ 85706}

\author[0000-0003-2599-7524]{Adam Smercina}
\affiliation{Department of Astronomy, University of Washington, Box 351580, U.W., Seattle, WA 98195-1580, USA}

\author[0000-0002-9912-6046]{Petia Yanchulova Merica-Jones}
\affiliation{Space Telescope Science Institute, 3700 San Martin Dr., Baltimore, MD 21218, USA}

\author[0000-0002-6442-6030]{Daniel R. Weisz}
\affiliation{Department of Astronomy, University of California, Berkeley, Berkeley, CA, 94720, USA}

\author[0000-0002-5564-9873]{Eric F. Bell}
\affiliation{Department of Astronomy, University of Michigan, 1085 S. University Ave, Ann Arbor, MI 48109-1107}

\author[0000-0003-0394-8377]{Karoline M. Gilbert}
\affiliation{Space Telescope Science Institute, 3700 San Martin Dr., Baltimore, MD 21218, USA}
\affiliation{The William H. Miller III Department of Physics \& Astronomy, Bloomberg Center for Physics and Astronomy, Johns Hopkins University, 3400 N. Charles Street, Baltimore, MD 21218, USA}

\author[0000-0002-6301-3269]{L\'eo Girardi}
\affiliation{Osservatorio Astronomico di Padova -- INAF, 
  Vicolo dell'Osservatorio 5, I-35122 Padova, Italy}

\author[0000-0001-5340-6774]{Karl Gordon}
\affiliation{Space Telescope Science Institute, 3700 San Martin Dr., Baltimore, MD 21218, USA}

\author[0000-0001-8867-4234]{Puragra Guhathakurta}
\affiliation{University of California Santa Cruz, 1156 High Street, Santa Cruz, CA, 95064}

\author[0000-0001-6421-0953]{L. Clifton Johnson}
\affiliation{Center for Interdisciplinary Exploration and Research in Astrophysics (CIERA) and Department of Physics and Astronomy, Northwestern University, 1800 Sherman Ave., Evanston, IL 60201, USA}

\author[0000-0003-3234-7247]{Tod R. Lauer}
\affiliation{NOAO, 950 North Cherry Avenue, Tucson, AZ 85721}

\author[0000-0003-0248-5470]{Anil Seth}
\affiliation{Department of Astronomy, University of Utah}

\author[0000-0003-0605-8732]{Evan Skillman}
\affiliation{Department of Astronomy, University of Minnesota, 116 Church SE, Minneapolis, MN 55455}


\begin{abstract}

We present the final legacy version of stellar photometry for the Panchromatic Hubble Andromeda Treasury (PHAT) survey. We have reprocessed all of the {\it Hubble Space Telescope} (HST) {\it
  Wide Field Camera 3} (WFC3) and {\it Advanced Camera for Surveys}
(ACS) near ultraviolet (F275W, F336W), optical (F475W, F814W), and
near infrared (F110W, F160W) imaging from the PHAT survey using an improved method that optimized the survey depth and chip gap coverage by including all overlapping exposures in all bands in the photometry.  An additional improvement was gained through the use of charge transfer efficiency (CTE) corrected input images, which provide more complete star finding as well as more reliable photometry for the NUV bands, which had no CTE correction in the previous version of the PHAT photometry.  While this method requires significantly more computing resources and time than earlier versions where the photometry was performed on individual pointings, it results in smaller systematic instrumental completeness variations as demonstrated by cleaner maps in stellar density, and it results in optimal constraints on stellar fluxes in all bands from the survey data.  Our resulting catalog has 138 million stars, 18\% more than the previous catalog, with lower density regions gaining as much as 40\% more stars.  The new catalog produces nearly seamless population maps which show relatively well-mixed distributions for populations associated with ages older than 1-2 Gyr, and highly structured distributions for the younger populations.

\end{abstract}

\section{Introduction}

Resolved stellar photometry of nearby galaxies has provided some of the best constraints to date on fundamental astrophysical processes, such as the initial mass function \citep[e.g.,][]{massey2003,elmegreen2006,bastian2010,weisz2015}, the distance scale \citep[e.g.,][]{leavitt1912,riess1995,freedman2001,dalcanton2009,durbin2020}, the progenitor masses of supernovae \citep[e.g.,][]{badenes2009,gogarten2009,williams2018,maund2018,diaz2018,koplitz2021}, and the history of star formation in galaxies \citep[][and many others]{gallart1999,holtzman1999,dolphin2000b,williams2002,dohm-palmer2002,harris2004,dolphin2005,weisz2011,weisz2014,williams2017,albers2019}.  While the large-scale detailed analysis of Galactic stellar populations made possible by large surveys such as Gaia \citep{gaia2016} is now revolutionizing our ability to test Galactic structure and evolution models \citep[e.g.,][]{helmi2018,belokurov2018,clarke2022} the ability to compare these to other massive spiral galaxies is critical to putting the Milky Way in context.

As the closest large spiral to our own, M31 is possibly the best
galaxy to make such comparisons to the Milky Way.  Furthermore, along
with the Milky Way, M31 is similar in physical properties to the
galaxies that make up most of the stellar mass in the universe
\citep[massive disk-dominated galaxies of roughly solar
  metallicity;][]{driver2007,gallazzi2008}.  With all of the key
science applications of such large resolved stellar catalogs in mind,
we undertook the Panchromatic Hubble Andromeda Treasury (PHAT) survey, covering roughly one third of the high surface brightness M31 disk with 414 HST pointings \citep{dalcanton2012}.

In \citet{williams2014} (hereafter \citetalias{williams2014}), we released a 6-band
photometry catalog covering the entire PHAT coverage, including 117
million detections.  Please see that paper and \citet{dalcanton2012}
for in-depth discussions of the history of stellar photometry in M31
and the PHAT survey motivations and strategy.  In \citet{williams2018b},
we released the same catalog on the NOAO datalab database server to
allow for easy searching and facilitate the use of the photometry from
each of the individual exposures for community time domain studies.
The catalog has since been applied to studies of merger history \citep{williams2015, hammer2018, dsouza2018}, chemical evolution \citep{telford2019}, dust
composition \citep{dalcanton2015,gordon2016,whitworth2019}, star formation rate
diagnostics \citep{lewis2017}, and kinematics from spectroscopic followup \citep{dorman2015,quirk2019,quirk2020}.  However, there are some known systematic
uncertainties and detector dependent depth issues with this first
complete catalog \citepalias{williams2014}.

The largest contributors to these issues were the ACS chip gaps and
the CTE corrections.  \citetalias{williams2014} measured photometry separately for each pointing of the survey.  Each pointing had dithers that covered the UVIS chip gap, but not the ACS chip gap, so the photometry for each pointing  was missing optical photometry in the ACS chip gaps.  This optical photometry was then taken from the photometry of the neighboring pointings, which had ACS data that covered the chip gap, which was matched to the other bands when the catalogs were merged together. As a result, \citetalias{williams2014}
found that the ACS chip gaps did not have the same photometric
sensitivity as the rest of the survey, especially in crowded regions.
The difference was due to the IR data not having the ACS data as a
prior in these regions during photometry.  It appeared that if we
could include the overlapping ACS images from the neighboring
pointings when measuring the photometry, so that such catalog-level merging was nnot necessary, then the sensitivity would remain
consistent across the chip gaps in the crowded regions.  At the same
time, \citetalias{williams2014} noted that there were systematic errors related to the
distance from the ACS chip gap that suggested the post-photometry CTE
corrections were sub-optimal. Finally, \citetalias{williams2014} used the
\citet{anderson2006} point spread functions (PSFs) for photometry, and
through further testing, we have found that while those PSFs appear to
work better than TinyTim \citep{krist2011} for astrometry, the TinyTim PSFs
perform better for photometry.

Herein we describe and supply the latest, and final, photometry
catalog for the PHAT survey that will be supplied by the original
survey team.  This catalog represents our best attempt to optimize the
depth of the resolved stellar photometry possible with the imaging
data.  We have addressed all of the shortcomings of the previous
catalog by including all of the overlapping exposures in all
photometric measurements, using CTE-corrected images, and fitting with
the TinyTim PSFs.  In Section 2 we describe details of the data
processing, focusing on differences between this processing and that
of \citetalias{williams2014}.  In Section 3, we present our results as a series of stellar
density and color-magnitude diagrams that help to assess the
photometric quality.  In Section 4, we discuss some direct comparisons
between the legacy catalog and \citetalias{williams2014}, and Section 5 provides a summary
of the work.  Throughout the paper we assume a distance to M31 of 770
kpc \citep{mcconnachie2005} for all angular conversions to physical
distance.
 
\section{Data}

The PHAT survey imaging data are described already in detail in \citetalias{williams2014}, and we provide only a brief description here for convenience. Images were obtained from July 12,
2010 to October 12, 2013 using the Advanced Camera for Surveys (ACS)
Wide Field Channel (WFC), the Wide Field Camera 3 (WFC3) IR (infrared)
channel, and the WFC3 UVIS (Ultraviolet-Optical) channel.  The
observing strategy is described in detail in \citet{dalcanton2012}.
In brief, we performed 414 2-orbit visits.  Each visit was matched by
another visit with the telescope rotated at 180 degrees, so that each
location in the survey footprint was covered by the WFC3/IR,
WFC3/UVIS, and ACS/WFC cameras.  The different orientations were
schedulable 6 months apart from one another, so that the WFC3 data and
ACS observations of each region were separated by 6 months. See \citetalias{williams2014}
for a full table of observations.

\begin{figure}
\centering
\includegraphics[width=5.0in]{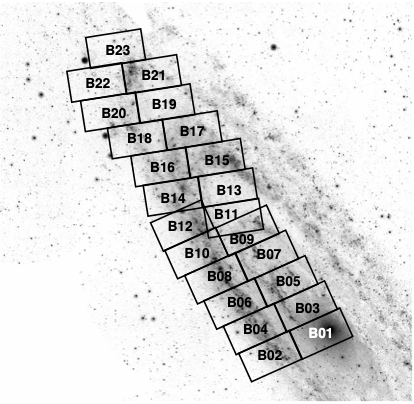}
    \caption{The location of the 23 bricks of the PHAT survey.  Each of these bricks consists of 18 HST pointings; each of which contains data in all 3 HST cameras. A full survey observing strategy description is provided in W14.}
    \label{fig:bricks}
\end{figure}

Each survey region is described by two identifiers, a ``brick'' number
and a ``field'' number. ``Bricks'' are rectangular areas of
$\sim$6$'{\times}12'$ that correspond to a 3$\times$6 array of WFC3/IR
footprints.  The bricks are numbered, 1-23, starting from the brick at
the nucleus, and counting west to east, south to north.  Odd numbered
bricks move out along M31's major axis, and even numbered bricks
follow the eastern side of the odd-numbered bricks, as shown in
Figure~\ref{fig:bricks}.  Within each of the 23 bricks, there are 18
``fields'' that each correspond to an IR footprint.  These fields are
numbered, 1-18, from the northeast corner, counting east to west,
north to south (see Figure~\ref{fig:bricks}, reproduced here from
\citetalias{williams2014} for convenience).

There were a few observations during which there were technical issues that required us to repeat the observation, complicating the
bookkeeping of observations.  Because of our parallel imaging
strategy, these repeat observations always affected 2 fields in a
brick.  These were for the pointings of Brick 01, Fields 3 and 6;
Brick 02, Fields 7 and 10; Brick 03, Fields 12 and 15; Brick 07,
Fields 2 and 5; Brick 10, Fields 14 and 17; Brick 16, Fields 11 and 14.
In the end, full coverage was obtained with the same exposure time as
proposed, so that these bookkeeping issues did not impact the final
coverage or depth of the survey.

%As described in \citetalias{williams2014}, there were systematic uncertainties that were correlated with the locations of the stars on the ACS detectors.  The largest contributors to these issues were the ACS chip gaps and
%the CTE corrections.  By reducing each pointing individually, \citetalias{williams2014}
%found that the ACS chip gaps did not have the same photometric
%sensitivity as the rest of the survey, especially in crowded regions.
%The difference was due to the IR data not having the ACS data as a
%prior in these regions during photometry.  It appeared that if we
%could include the overlapping ACS images from the neighboring
%pointings to cover the chip gap, then the sensitivity would remain
%consistent across the chip gaps in the crowded regions.  At the same
%time, \citetalias{williams2014} noted that there were systematic errors related to the
%distance from the ACS chip gap that suggested the post-photometry CTE
%corrections were sub-optimal. Additionally, no CTE corrections were
%made to the WFC3 data at all.  Finally, \citetalias{williams2014} used the
%\citet{anderson2006} point spread functions (PSFs) for photometry, and
%through further testing, we have found that while those PSFs appear to
%work better than TinyTim \citep{krist2011} for astrometry, the TinyTim
%PSFs perform better for photometry.

\section{Photometry}

All of the photometry presented here was performed using the same software packages as described in \citetalias{williams2014}.  Namely, we used the DOLPHOT software package \citep{dolphin2000b,dolphin2005,dolphin2016} to perform point spread function fitting on all of the individual aligned frames.  All of the photometry parameters for DOLPHOT were the same as \citetalias{williams2014}, except that we set the PSF library to be those from TinyTim \citep{krist2011} instead of those from \citet{anderson2006}, since this setting resulted in lower spatial systematics in tests since \citetalias{williams2014}.  Thus for the technical details for our use of DOLPHOT, the reader can see the full description in \citetalias{williams2014}.  Here we describe in detail only the differences between those measurements and this legacy photometry.

The most significant change for this legacy photometry was how we divided the data for each photometry run.  We divided the data into half-bricks, totaling 46 sets of images.  In each of these 46 photometry runs, we included all of the exposures that overlapped at all with the half-brick region.  Thus, neighboring half-brick sets had many exposures in common, namely, all of the exposures that overlapped with the boundary, providing consistent photometry along all of the edges, as well as the deepest possible photometry inside of the half-brick boundaries.  

Another significant improvement was that we used only CTE-corrected exposures ({\tt flc}) for all UVIS and ACS imaging.  This improvement removed the need to adopt catalog-level CTE corrections based on Y-pixel position for the ACS data,  as was done to the ACS data in \citetalias{williams2014}.  Furthermore, \citetalias{williams2014} performed no CTE corrections at all for the UVIS photometry, and by using CTE-corrected UVIS data for this legacy reduction, we have now taken CTE into account.  

\begin{figure}
    \centering
    \includegraphics[width=6.0in]{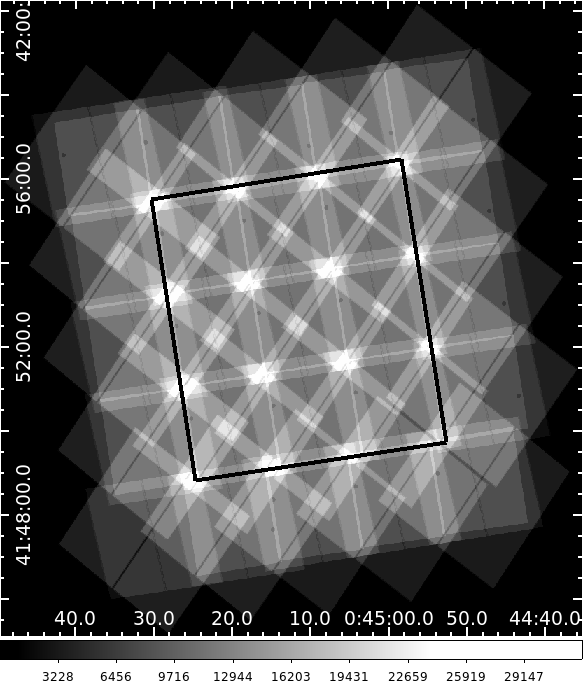}
    \caption{Exposure map of all exposures included in a typical half-brick run, where photometry was only performed on the area indicated withe the black box, but all exposures that had pixel falling within that area were included in the run, resulting in our seamless final catalog.}
    \label{fig:exposuremap}
\end{figure}

Finally, another slight modification to the \citetalias{williams2014} process was that we saved computational power by only searching for stars
and measuring their photometry if they were within $1\arcsec$ of the
half-brick boundary.  In this way, we generated 46 catalogs wherein
all of the stars were measured using every relevant PHAT exposure.  
%We
%included some overlap in these regions, allowing for consistency
%checks across the overlapping areas; 
However we did not include any
duplicate measurements from the small overlaps in the final catalogs.  Figure~\ref{fig:exposuremap} shows an example half-brick (eastern half of brick 15), where the outline of the half-brick is drawn on an exposure map of all of the exposures that overlap with that outline.

\section{Artificial Star Tests}

As in \citetalias{williams2014}, to assess the quality of our photometry as a function of magnitude in each band and as a function of the stellar density of location within the survey, we performed 250,099 artificial star tests (ASTs).  For each of these tests, a star of known brightness in all bands was added to all of the exposures that were included in the original photometry measurements.  Then the photometry routine, including source detection, is rerun on the data, and the results are checked for the detection and measurement of the artificial star that was added to the data.  

The first step in performing ASTs is to generate an input star catalog.  Since star finding is performed on the full stack of all exposures, including all bands, the most accurate characterization of the photometry will result from input spectral energy distributions (SEDs) that mimic those of single stars.  To generate such SEDs, we use the Bayesian Extinction and Stellar Tool \citep[BEAST;][]{gordon2016} software package.  This package contains functions that will produce a grid of stellar SEDs from current models. For our purposes, we applied the PARSEC and COLIBRI evolutionary models to create a grid of effective temperatures ($T_{eff}$), bolometric luminosity $L_{bol}$, and surface gravity log($g$) \citep[]{bressan2012,marigo2017}.  We then generate intrinsic SEDs using stellar atmosphere models of \citet{castelli2003} and TLUSTY \citep{lanz2003,lanz2007}.  

The BEAST adjusts these intrinsic SEDs using the distance modulus of the observed sample (in this case, the distance to M31), and it produces a set of SEDs for each model at each of a variety of extinction values,  and extinction law parameters described in \citet{gordon2016}.  Once it has a well-sampled grid of relevant models that cover the range of observed stellar brightnesses, it generates input positions for the stars so that a full sample of ASTs is included at each of a range of stellar densities.  This carefully-determined spatial distribution of the input lists provides a full characterization of the photometric bias, uncertainties, and completeness as a function of crowding across the observed area. 

\begin{figure}
    \centering
    \includegraphics[width=6.5in]{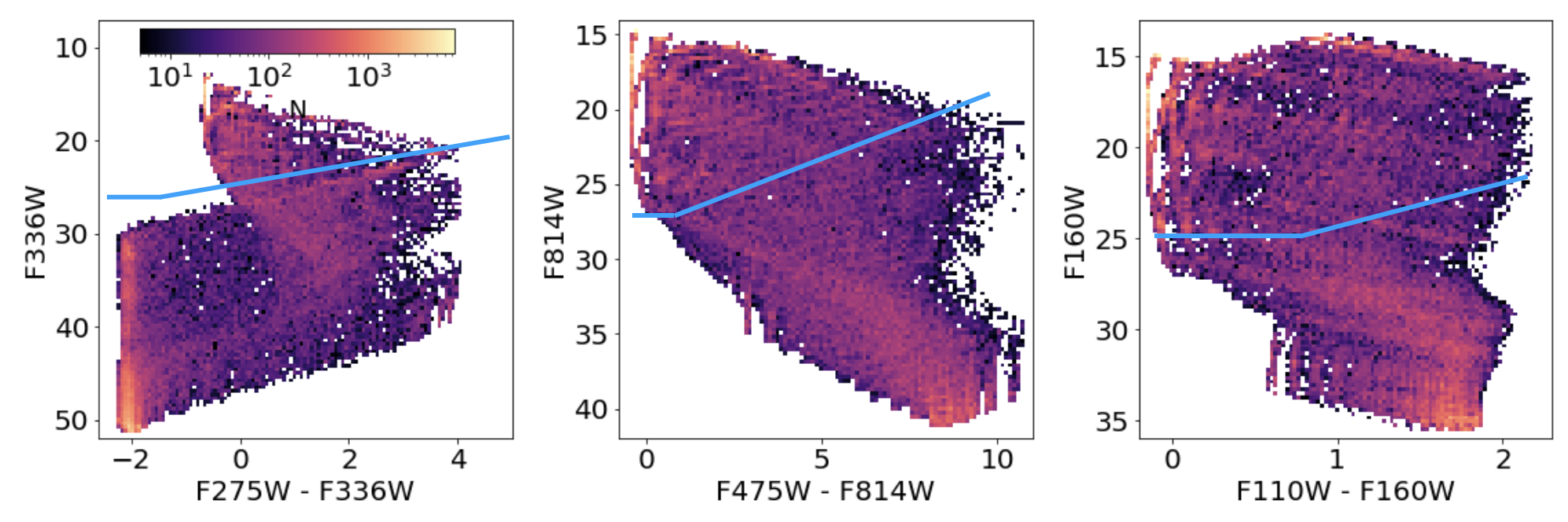}
    \caption{Input artificial star CMDs showing how the model SEDs cover the relevant color space. The color bar is in units of stars per CMD bin.  Blue lines mark the approximate depth of the observed photometry.  The models extend to much fainter fluxes than any detections in some bands due to very red colors of some models.  Specifically, some of the oldest ages, highest-metallicities, and high reddenings are potentially detectable in the IR, but extremely faint in the UV.  Further, the very faint blue feature in the UV bands is due to red leaks in the UV filters, which result in small fluxes for red stars that are not detectable in the UV bands.}
    \label{fig:input_asts}
\end{figure}

In Figure~\ref{fig:input_asts}, we show  color-magnitude diagrams (CMDs) from the input AST SEDs.  By including both a large grid of ages and metallicities in the model SEDs, as well as applying a range of extinction values, we are able to fully cover the range of observed fluxes in all bands with sufficient sampling.  In total, we sampled 1190112 model SEDs covering $-0.6<=$[Fe/H]$<=0.3$, $6<=$log(age)$<=10.13$ and $0<={\rm A_V}<=10$, with the sample repeated in multiple areas of the survey with stellar densities ranging from $\sim$0 to $\sim$30 arcsec$^{-2}$, where the density was calculated using only stars with 20$<$F814W$<$22. We did not include lower metallicity models because they were not necessary to provide good coverage of the flux ranges seen in the data.  The final input star list included 250099 ASTs within at least 1 magnitude of the range of observed fluxes.  This sampling provided all of the necessary data to perform a basic characterization of our data quality in each band as a function of magnitude and stellar density.

\section{Results}

We provide our 6-band photometry, along with the signal-to-noise and quality flag, in Table ~\ref{tab:photometry_table}.   This small sample shows the included columns, while the full catalog is only available electronically.  The full catalog contains measurements of 137,852,215 sources.  The quality flag is provided to show which bands had measurements whose quality parameters (sharpness, signal-to-noise, and crowding) passed our quality cuts.  In total 429,697 passed these cuts in F275W, 2,871,968 in F336W, 99,422,134 in F475W, 113,568,963 in F814W, 80,526,872 in F110W, and 69,774,534 in F160W.  Those interested in generating their own cuts from the DOLPHOT quality parameters can access the full photometry tables, 
%including all of these parameters for each measurement
available at MAST as a High Level Science Product via DOI\dataset[10.17909/T91S30]{\doi{10.17909/T91S30}}.\footnote{\url{https://archive.stsci.edu/hlsp/phat/}}

\subsection{Color-Magnitude Diagrams}

\begin{figure}
    \centering
    \includegraphics{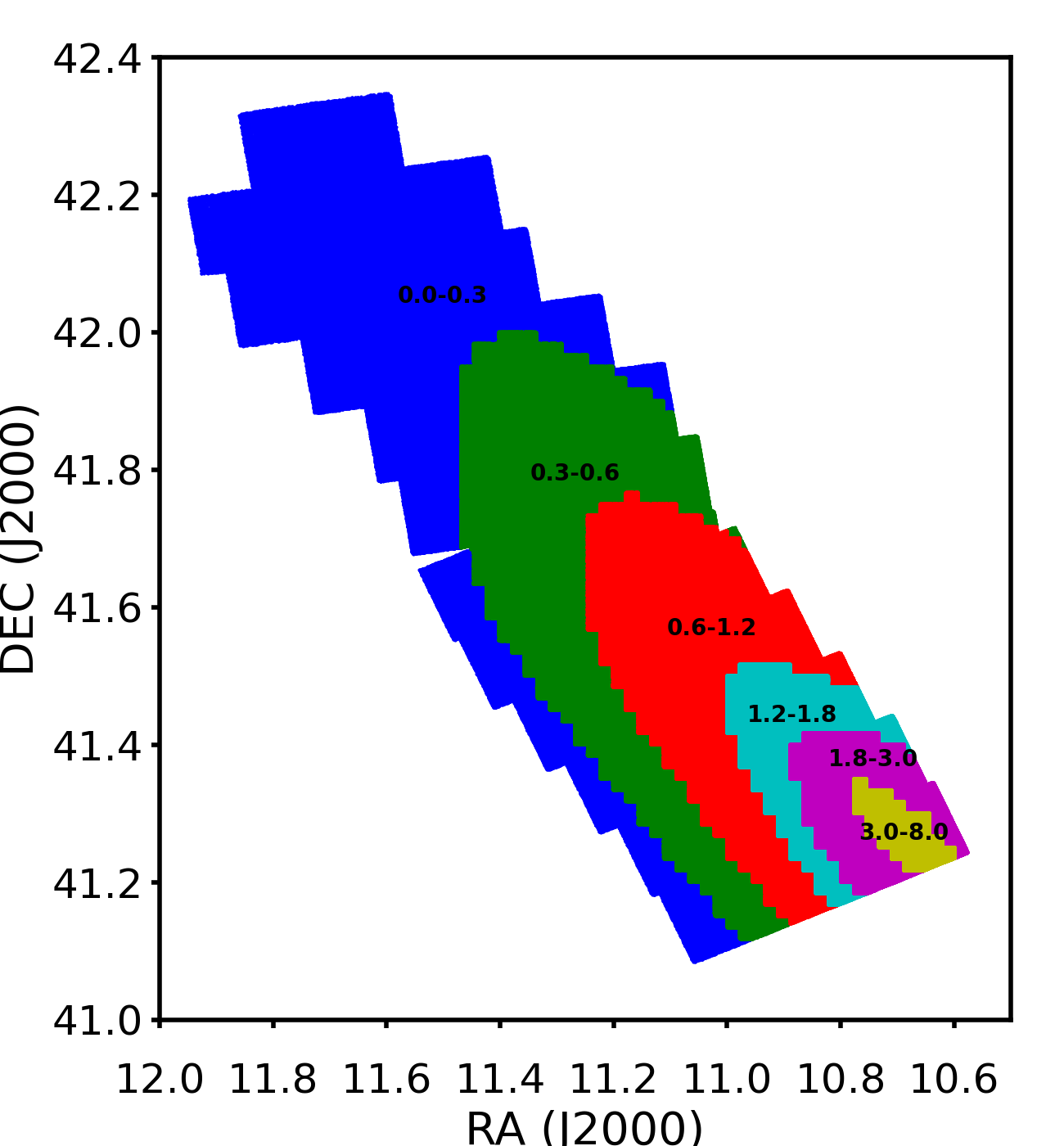}
    \caption{Map of the six levels of model stellar densities used to define the six regions that were used to make the different CMDs shown in Figures 11–18. Numbers refer to the log of the stellar density of red giant branch stars. The numbers are relative and determined from a model disk, making the normalization irrelevant. The areas covered by the defined regions are 770, 481 349, 113, 106, and 31 arcmin2 from lowest to highest stellar density, respectively. We divided our results in this way to account for the changing completeness limits with stellar density in the same way as W14 for easy comparison.}
    \label{fig:density_map}
\end{figure}

\begin{figure}
    \centering
    \includegraphics[width=4.0in]{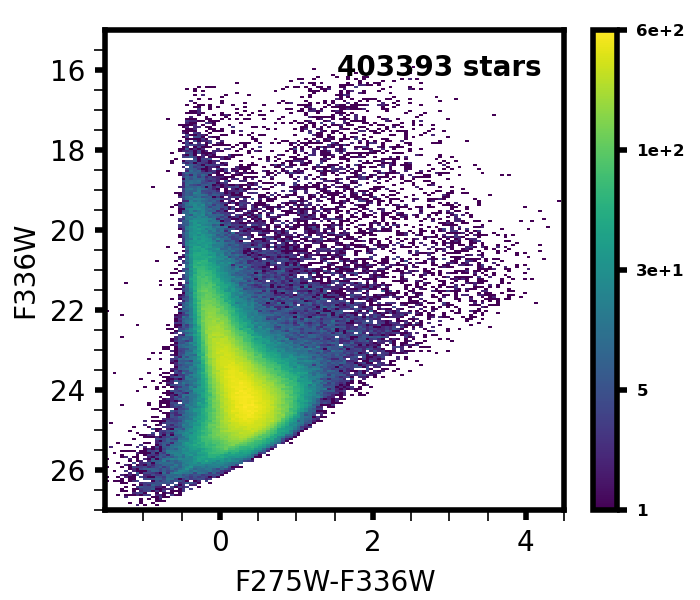}
    \caption{UV CMD for the entire survey.  The color bar is in units of stars per CMD bin.  The photometry extends slightly deeper ($\sim$0.2 mag) than W14 due to the inclusion of overlapping imaging when performing the photometry.}
    \label{fig:uv_cmd}
\end{figure}

\begin{figure}
    \centering
    \includegraphics[width=6.0in]{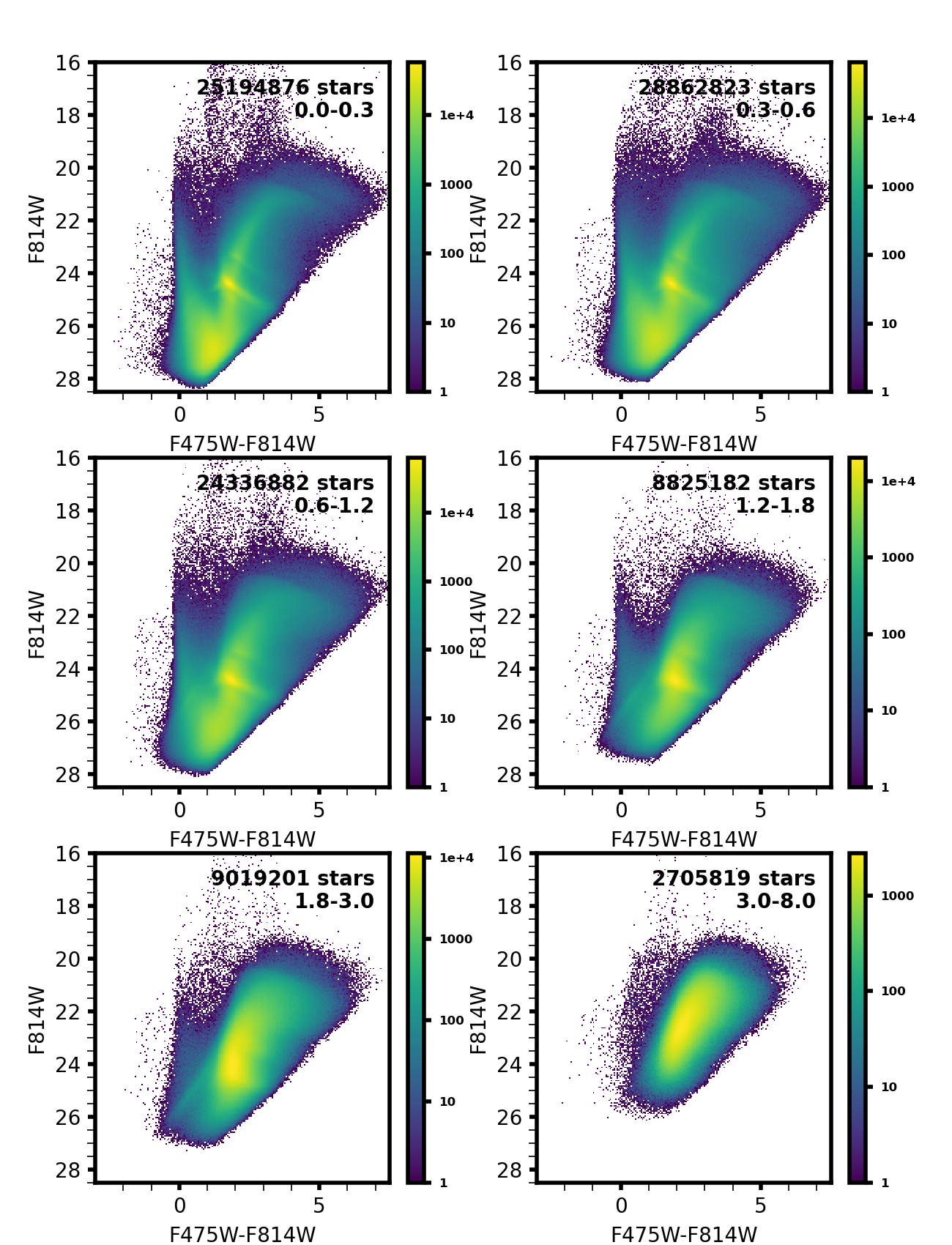}
    \caption{Optical CMDs for the 6 colored regions shown in Figure~\ref{fig:density_map}.  The color bar is in units of stars per CMD bin. The rows go from lower to higher relative stellar density to show the decrease in photometric quality as stellar density increases. The photometry extends slightly deeper ($\sim$0.2 mag) than W14 at all but the highest stellar density region.}
    \label{fig:opt_panels_cmd}
\end{figure}

\begin{figure}
    \centering
    \includegraphics[width=6.0in]{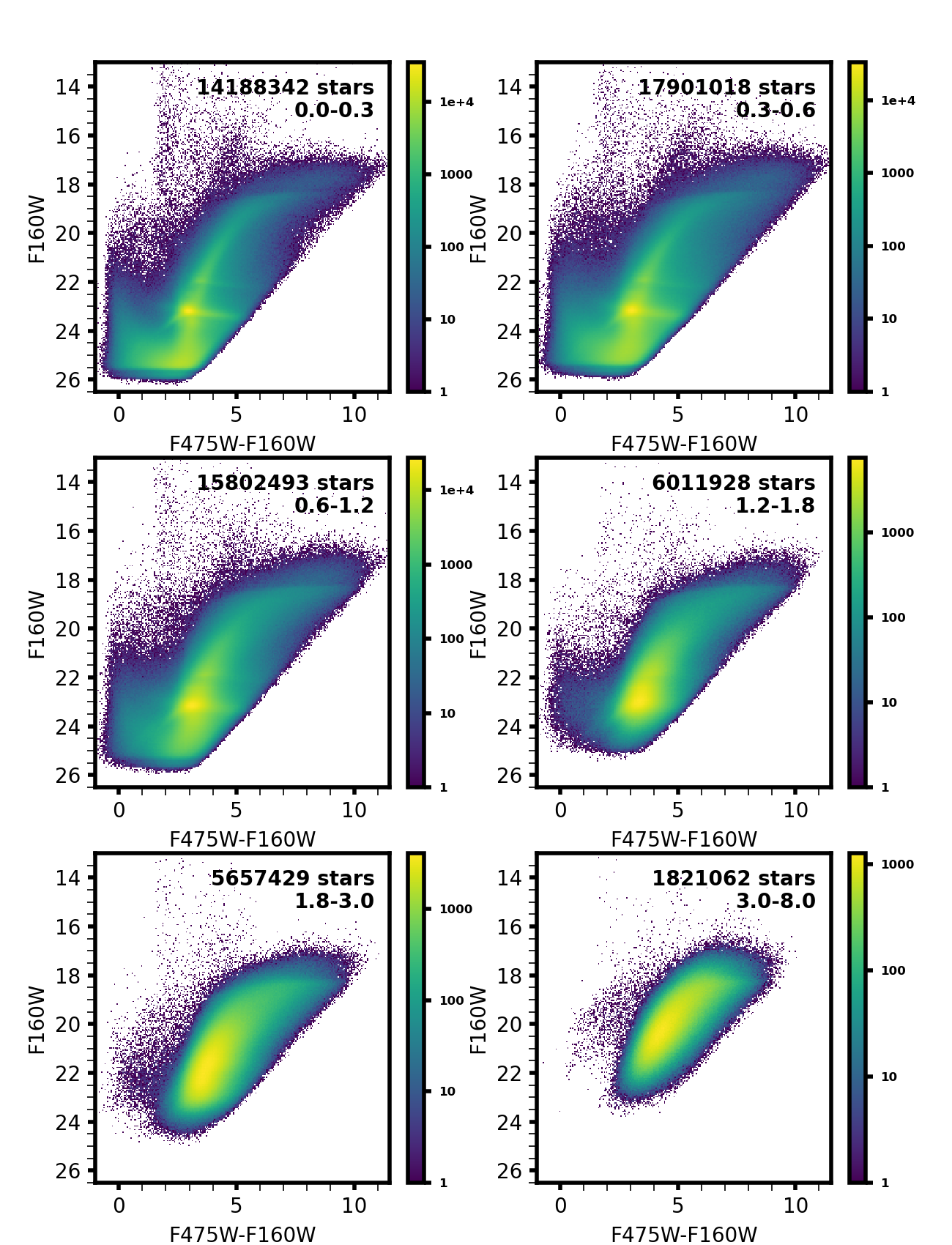}
    \caption{Optical-IR CMDs for the 6 colored regions shown in Figure~\ref{fig:density_map}.  The color bar is in units of stars per CMD bin. The rows go from lower to higher stellar density to show the decrease in photometric quality as stellar density increases. The photometry extends slightly deeper than W14 at all but the highest stellar density region.}
    \label{fig:optir_panels_cmd}
\end{figure}

\begin{figure}
    \centering
    \includegraphics[width=6.0in]{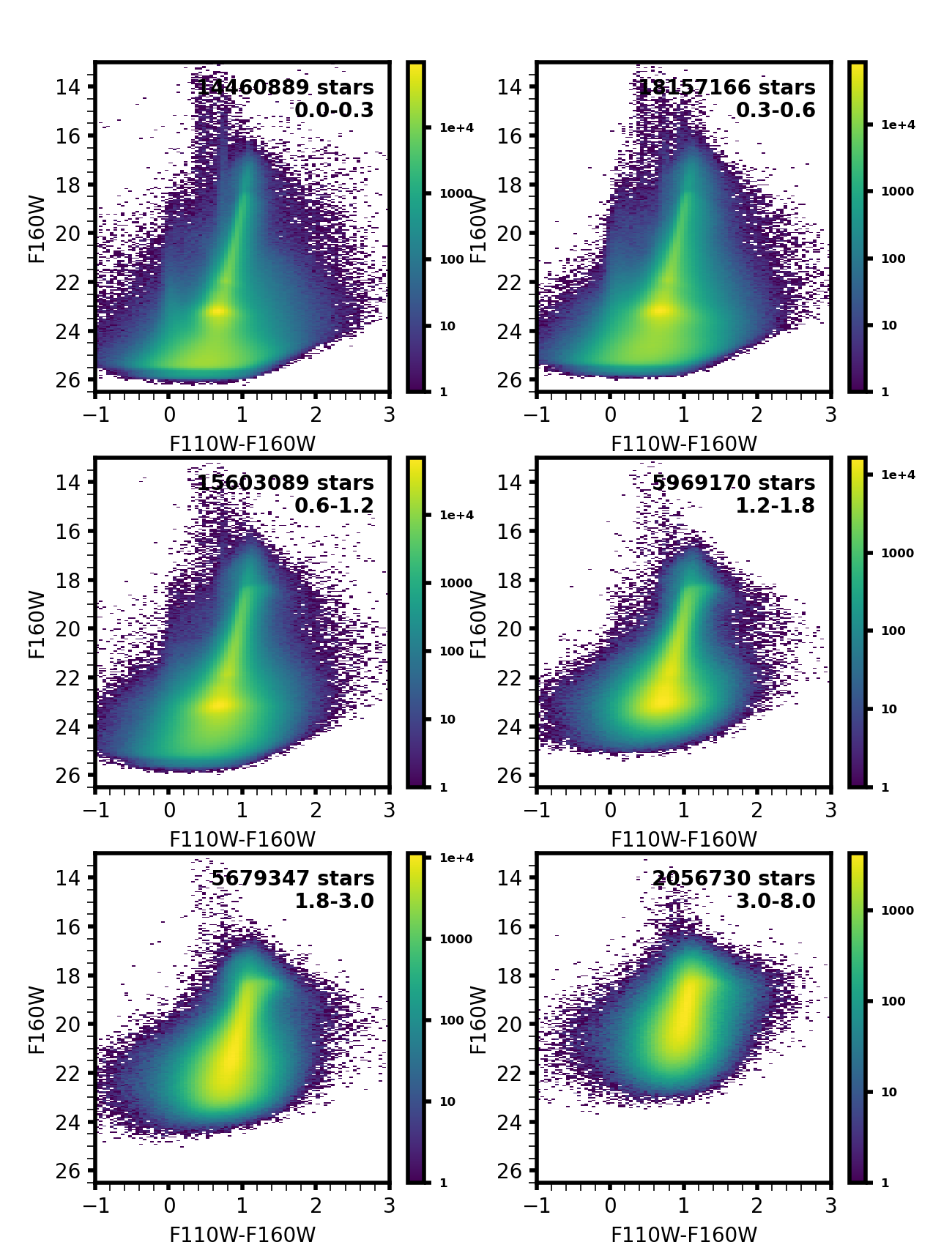}
    \caption{IR CMDs for the 6 colored regions shown in Figure~\ref{fig:density_map}.  The color bar is in units of stars per CMD bin. The rows go from lower to higher stellar density to show the decrease in photometric quality as stellar density increases.  }
    \label{fig:ir_panels_cmd}
\end{figure}

We also provide CMDs of the photometry analagous to those in W14. Figure~\ref{fig:uv_cmd} shows the UV CMD of the entire survey, as the UV imaging was not crowding limited.  Multiple depths are still visible in the UV CMD due to the varying amount of overlap between the pointings.  Figure~\ref{fig:density_map} shows the survey broken down into 6 regions by their relative  stellar density range as determined from a disk model.  Thus this map represents the amount of crowding in the data as a function of position in the survey.  Figures~\ref{fig:opt_panels_cmd}, \ref{fig:optir_panels_cmd}, and \ref{fig:ir_panels_cmd} show CMDs of the survey photometry in each of these regions, so that the effects of crowding on our photometry in these bands can be seen.  The CMDs of the higher stellar density regions have features that are not as sharp and do not extend as faint as in the CMDs of the lower stellar density regions.

\subsection{Comparisons to W14}

The latest and final resolved stellar photometry catalog from our
team, reported in this paper, is qualitatively similar to that
released in \citetalias{williams2014}.  However, it does have systematic differences at the
$<0.1$ mag level, and it has improved spatial uniformity due to the
treatment of the ACS chip gaps.  

\begin{figure}
    \centering
    \includegraphics[width=6.0in]{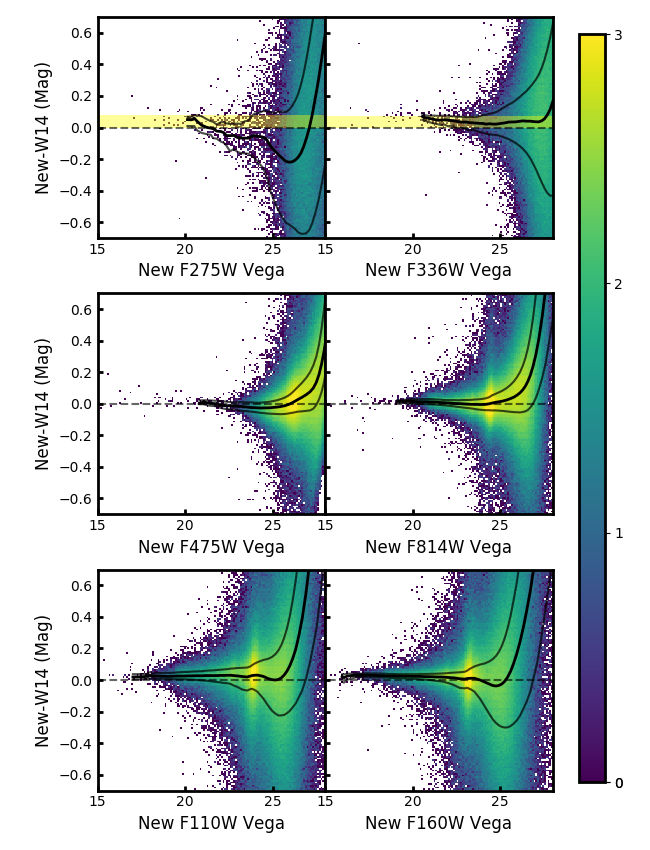}
    \caption{Example star-by-star comparison for all 6 bands, taken from Brick 8 of the survey.  The color bar is in units of log of the number of stars per $\Delta$mag-mag bin. The median of the residual is shown by the thick line, while the interquartile range is shown with the thin lines.  The residuals show general consistency between the measurements.  The UV bands show significant offsets, which is expected from updates to calibrations and CTE corrections, indicated by the yellow shading. The IR bands show slightly fainter new photometry, likely partially due to better deblending.  Furthermore, all bands show systematic bias to fainter magnitudes at the faint end, as expected from improved deblending.  }
    \label{fig:matched_cats1}
\end{figure}

\begin{figure}
    \centering
    \includegraphics[width=6.0in]{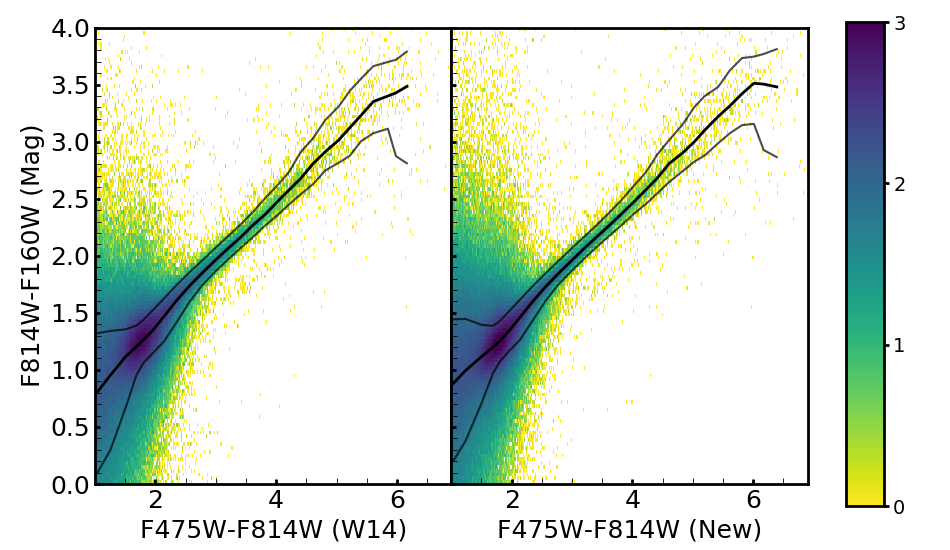}
    \caption{Color-color relations of red clump and red giant branch stars from matched regions of Brick 8 in the W14 catalog and our new catalog.  The color bar is in units of the log of the number of stars per color-color bin. The thick lines mark the median F814W-F160W color as a function of F475W-F814W color. While the results are clearly similar, the width of the 16th to 84th percentile of the F814W-F160W color as a function of F475W-F814W color, shown with the thinner lines, is slightly narrower (by a few hundredths of a magnitude) in the new photometry, confirming improved photometric precision over W14.}
    \label{fig:matched_cats_ccd1}
\end{figure}

The similarity of the overall
photometry is shown in Figures~\ref{fig:matched_cats1}-
\ref{fig:matched_cats_ccd1}.  Figure ~\ref{fig:matched_cats1} shows a star-by-star comparison of the
magnitudes in all 6 bands for a subset of stars matched between W14 and this work.   The
largest differences are in the UVIS bands, which is not surprising
since those flat fields and zero points have been revised since \citetalias{williams2014}\footnote{\url{https://www.stsci.edu/files/live/sites/www/files/home/hst/instrumentation/wfc3/documentation/instrument-science-reports-isrs/_documents/2017/WFC3-2017-07.pdf}}
and \citetalias{williams2014} did not correct for any CTE effects in these bands.  The zeropoint difference is indicated by the yellow shading in Figure~\ref{fig:matched_cats1}.  

Figure~\ref{fig:matched_cats_ccd1} compares the colors of red clump and red giant branch stars from matched areas in both catalogs.  The shape and size of the red giant branch in color-color space is also very similar to W14; however, quantitatively, the feature is slightly narrower (by a few hundredths of a magnitude) in our new reduction. The reduction in feature width is most likely due to smaller errors in photometry made possible by including more data at each star position. 

Overall, both figures show a similar amount of systematic offsets
between the current and previous measurements, all $<0.1$ mag.Thus,
this new photometry is an improvement over \citetalias{williams2014} and should supersede it
in any future analysis using 6-band resolved stellar photometry from PHAT.

\begin{figure}
    \centering
    \includegraphics[width=6.0in]{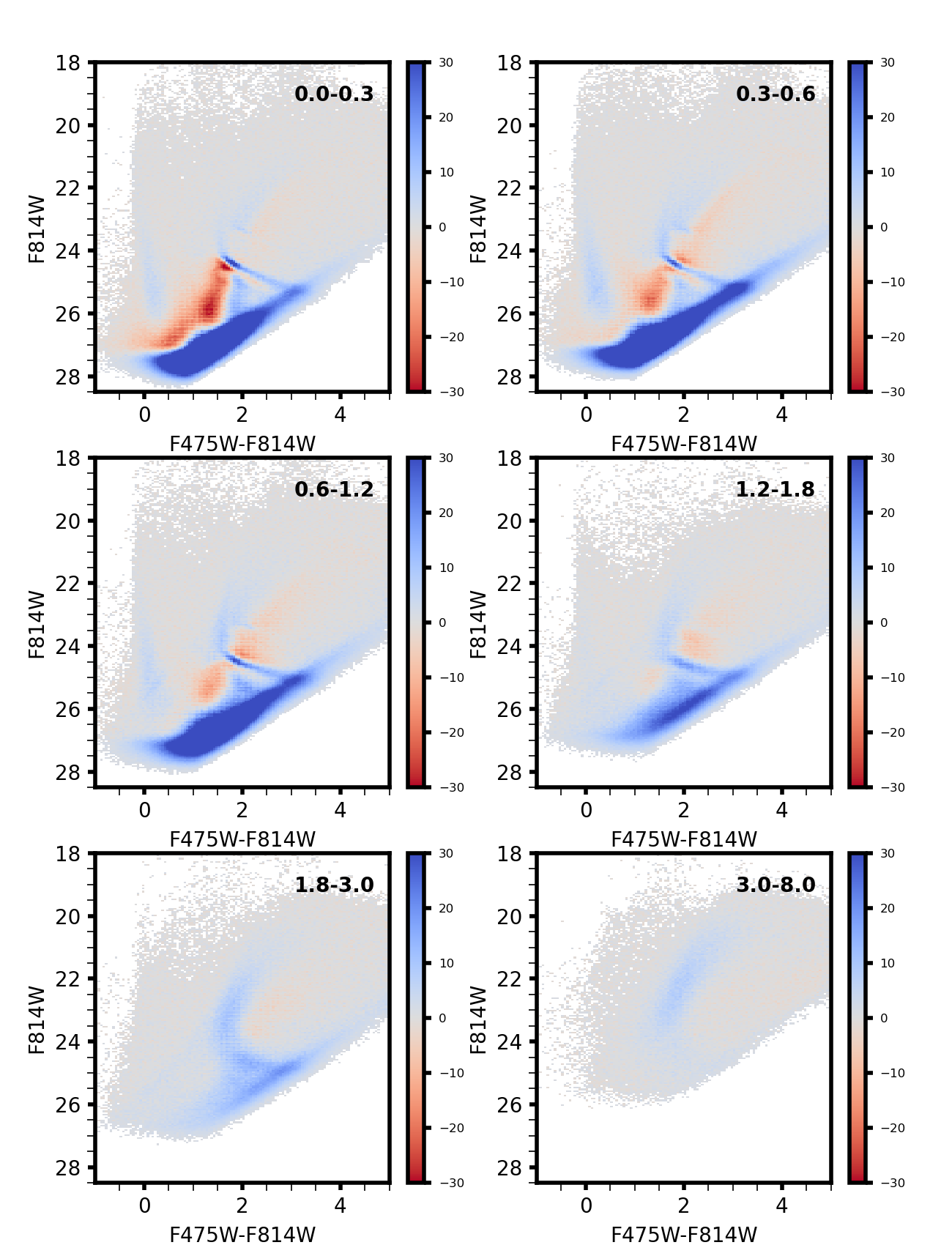}
    \caption{The new optical CMDs for the 6 colored regions shown in Figure~\ref{fig:density_map} with the W14 CMDs of the same regions subtracted away. The color bar indicates the significance of the difference between the two CMDs, calculated as the new CMD minus the old CMD divided by the square root of the sum of the two CMDs in each CMD bin.  The rows go from lower to higher stellar density to show the decrease in photometric quality as stellar density increases. The photometry extends slightly deeper than W14 at all but the highest stellar density region, as shown by the larger number of stars at fainter magnitudes.  Furthermore, the red clump feature  (the strip starting at F814W=24, F475W-F814W=1.5 and extending to the red) appears tighter when it is detected.}
    \label{fig:optdiff}
\end{figure}

\begin{figure}
    \centering
    \includegraphics[width=6.0in]{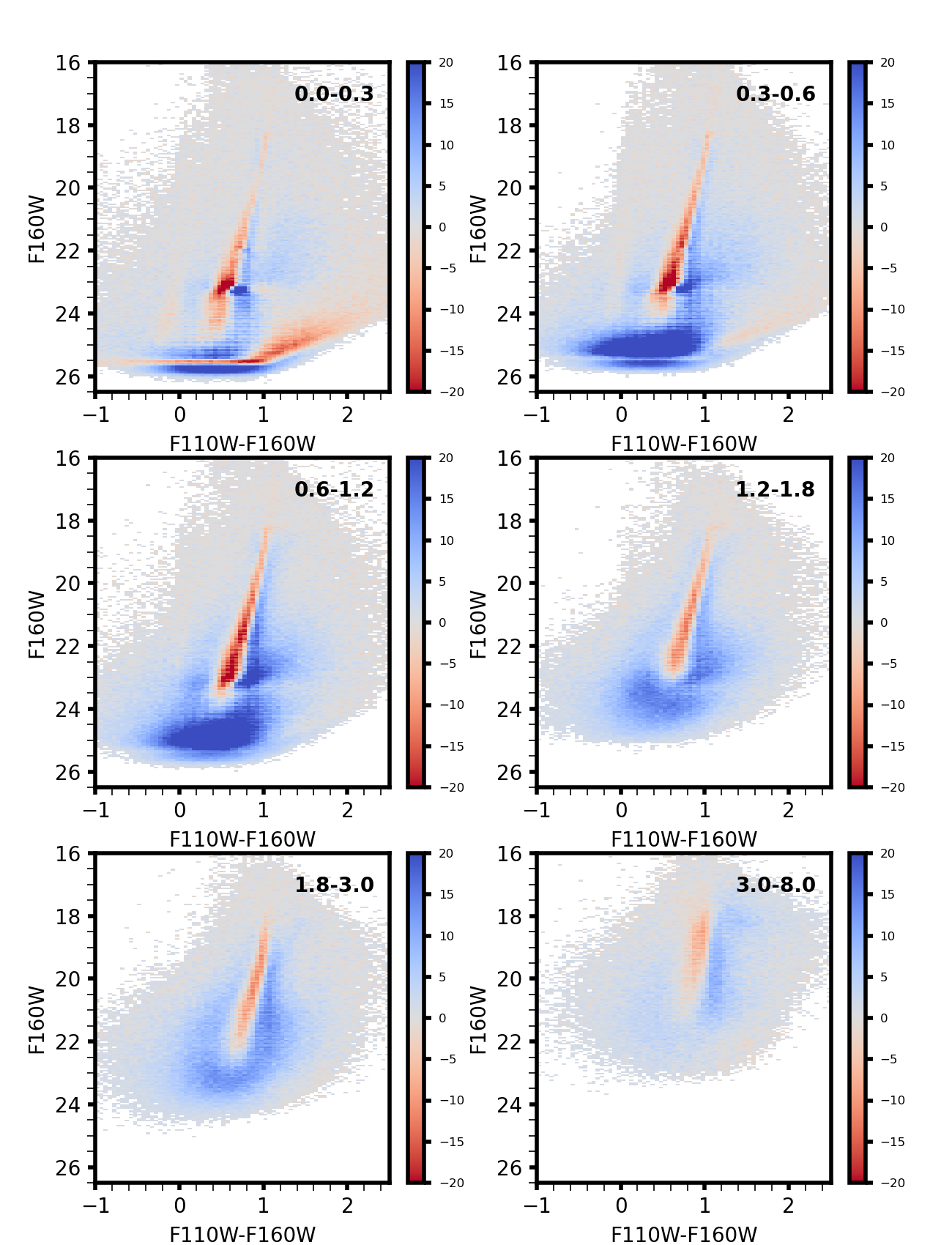}
    \caption{The new IR CMDs for the 6 colored regions shown in Figure~\ref{fig:density_map} with the W14 CMDs of the same regions subtracted away.  The color bar indicated the significance of of the difference as is calculated as in Figure~\ref{fig:optdiff}. The rows go from lower to higher stellar density to show the decrease in photometric quality as stellar density increases. The new photomety is slightly redder, likely due mostly to the  improved deblending from including all overlapping data in the star finding, as well as the different IR PSF.  Furthermore, the red clump is detected in the third aned fourth density bins (middle row) in the new photometry at more than 5$\sigma$ greater significance than in W14.}
    \label{fig:irdiff}
\end{figure}

To compare the overall photometric characteristics of the new catalog to the old, we subtracted the CMDs (new catalog - old catalog) of the 6 regions of the galaxy used to show the varying completeness.  These difference CMDs, shown in the optical and IR in Figures~\ref{fig:optdiff} and \ref{fig:irdiff} respectively,  highlight the features that changed between the two reductions.  Both figures show a higher number of faint stars in the new reductions for intermediate stellar densities.  The optical CMDs show narrower red clump feature (by $\sim$0.02 mag) at most stellar densities, and increased  red clump detections (by $\sim$10\% survey-wide).  The IR shows a clear systematic color difference where the new catalog measurements are $\sim$0.04 mag redder than those of the old catalog.  The new catalog also shows more faint detections at intermediate stellar densities than the old catalog.  The slightly redder IR colors are due to fainter F110W magnitudes, also seen in the star-by-star comparisons in Figure~\ref{fig:matched_cats1}.  These fainter F110W magnitudes are likely due in part to the different PSFs used here, and in part to improved deblending as a result of using the full stack of data for star finding.

\begin{figure}
    \centering
    \includegraphics[width=3.1in]{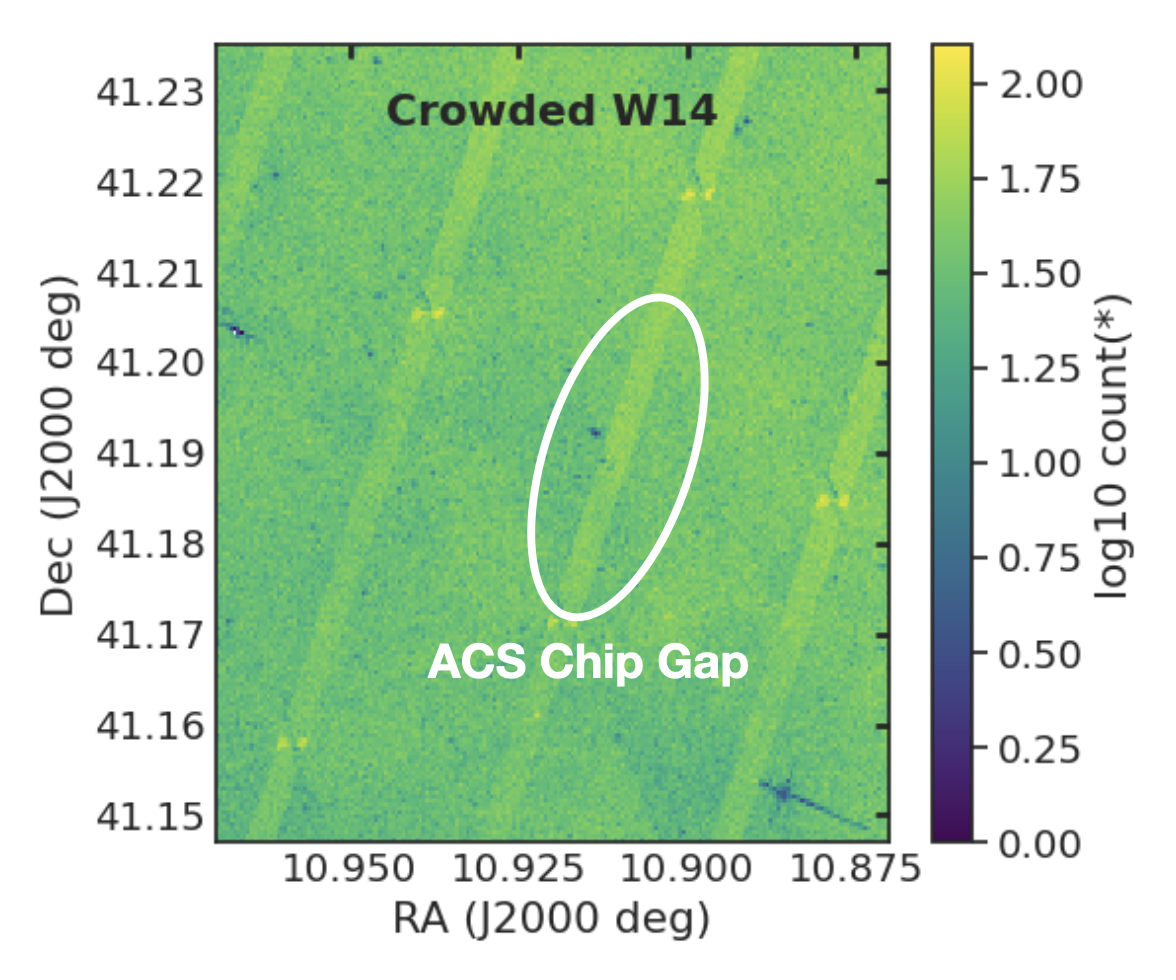}
    \includegraphics[width=3.1in]{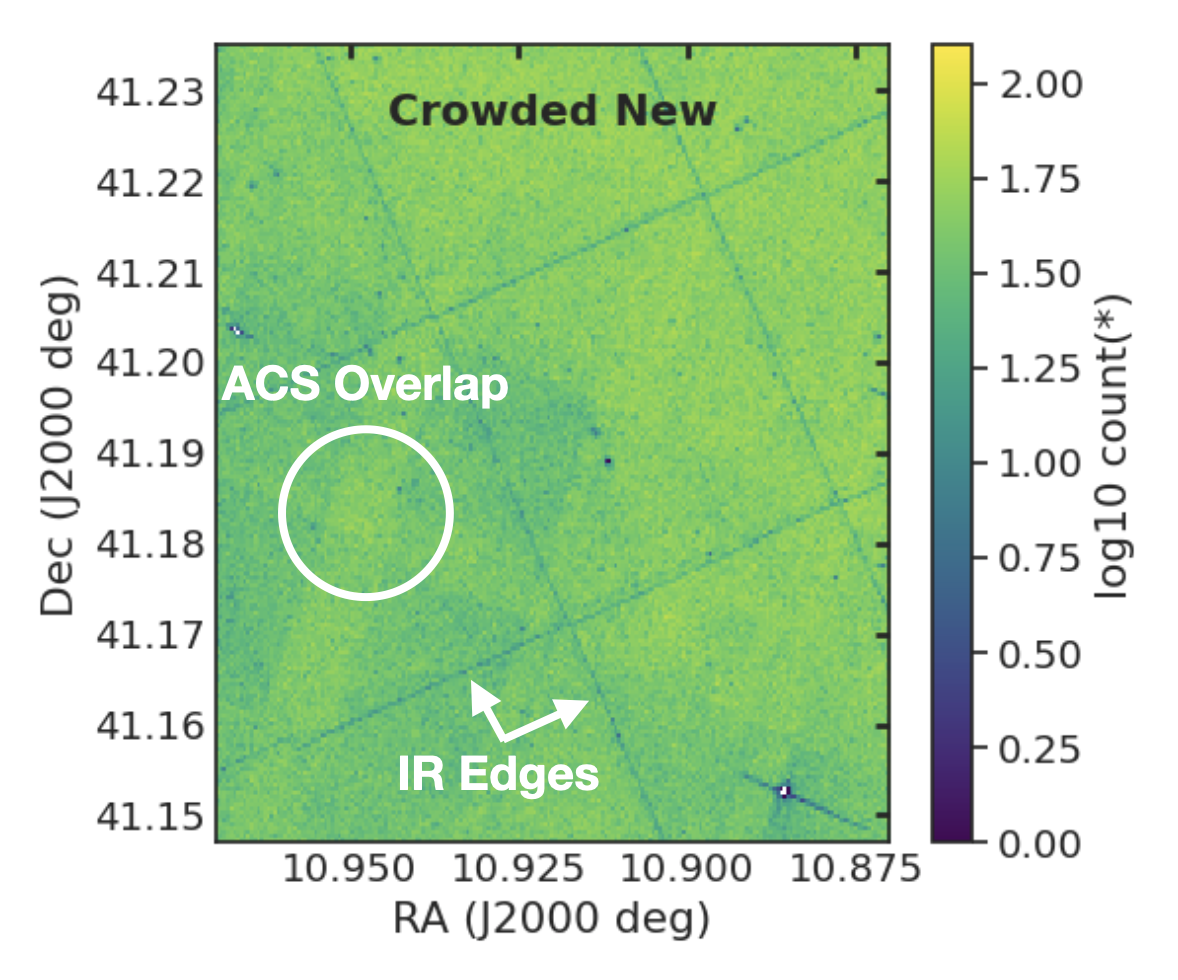}
    \includegraphics[width=3.1in]{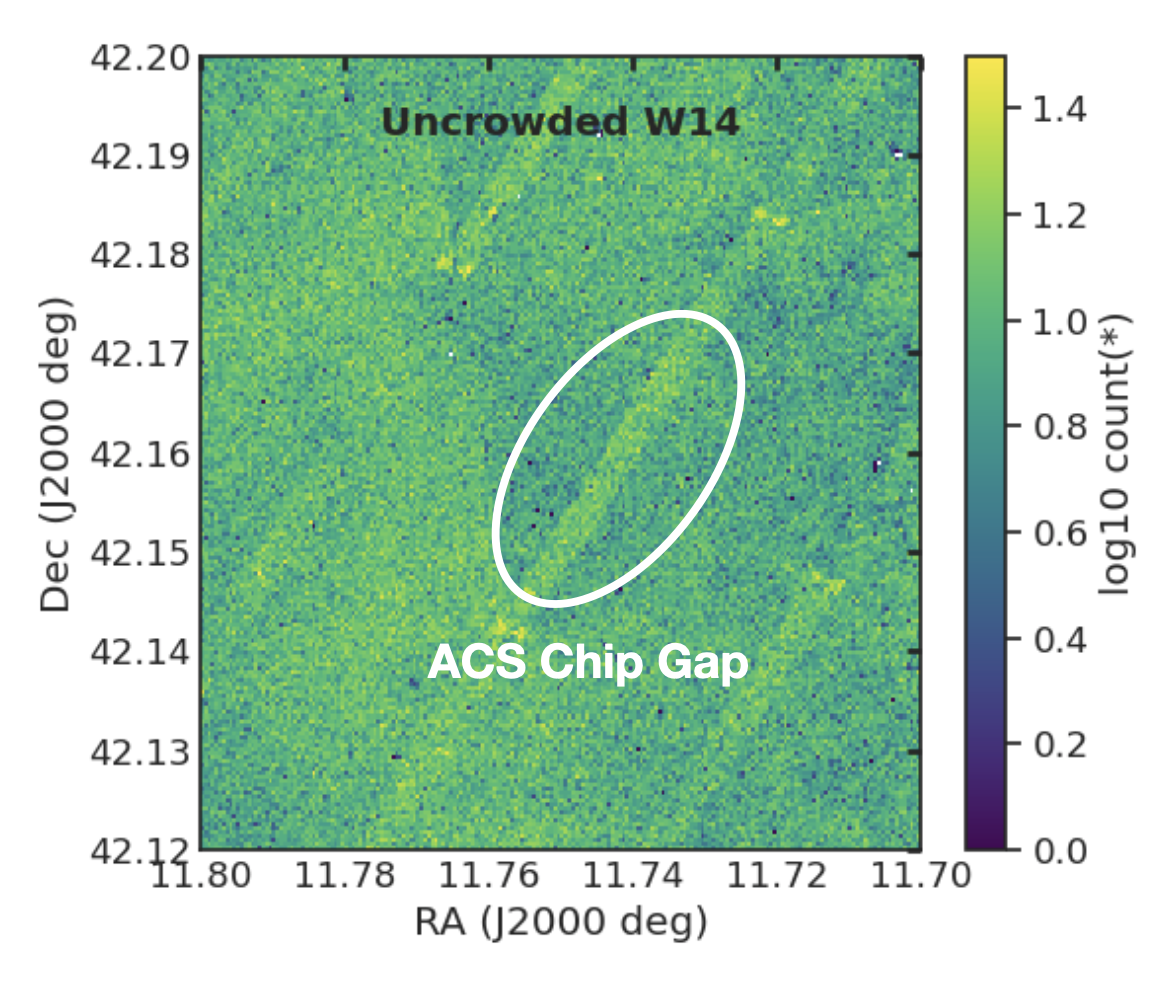}
    \includegraphics[width=3.1in]{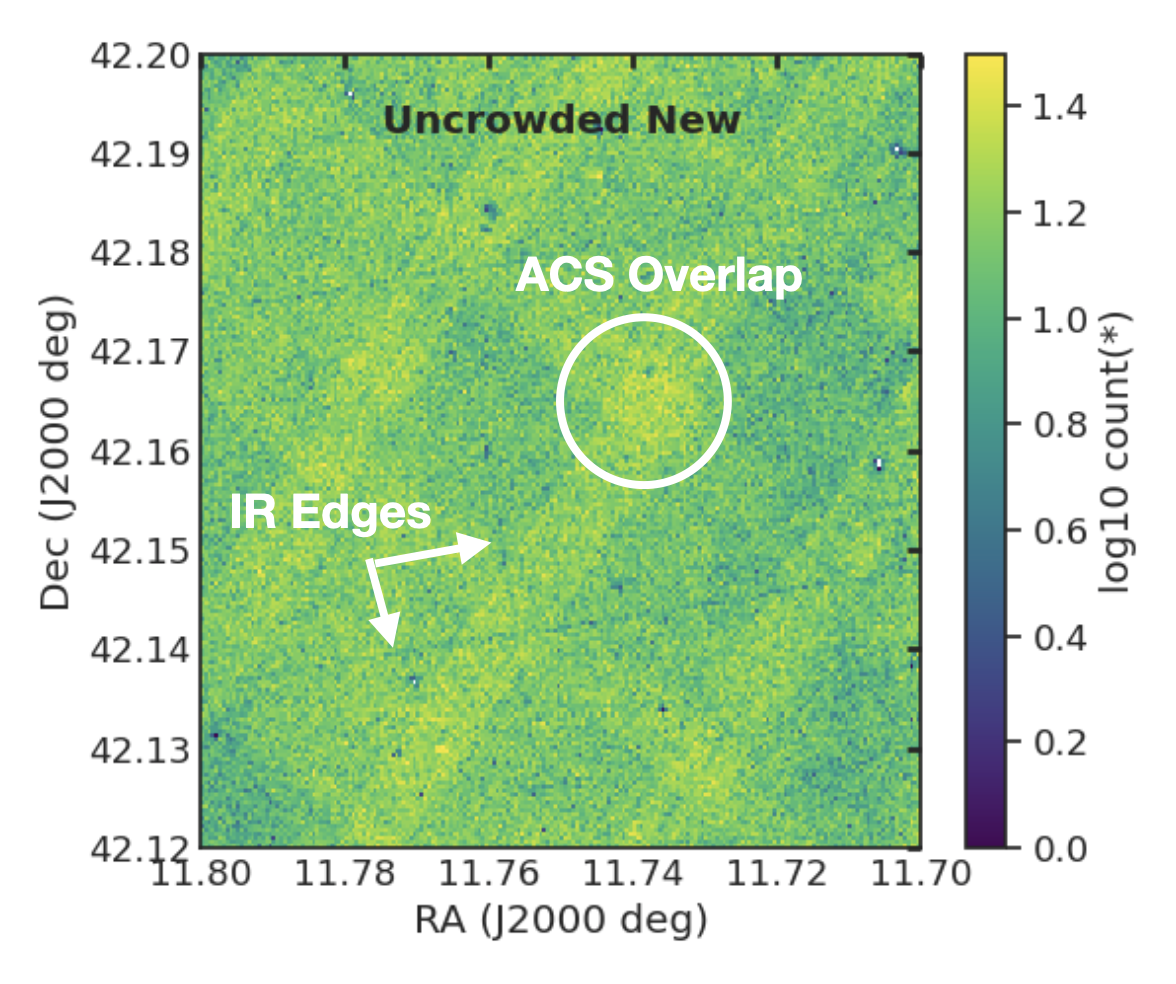}
    \caption{Comparison of the spatial distributions of a small section of the survey from W14 (left) and this work (right).  The color bar is in units of the log of the number of stars per spatial bin. {\it Top row:} A crowded region of the survey (Brick 2). There is improved depth, except at the IR detector edges (example labeled), and there is improved homogeneity in this legacy reduction. {\it Bottom row:} An uncrowded region (Brick 22). There is improved depth in the overlapping ACS frames (example labeled) and improved homogeneity in this legacy reduction. }
    \label{fig:spatial_comparison}
\end{figure}

The spatial distribution of the stars in the new catalog is also similar, but noticeably improved.  In Figure~\ref{fig:spatial_comparison}, we compare the stellar surface density in the \citetalias{williams2014} catalog with this work for the area of western halfs of Bricks 2 and 22.  The left panels show the stellar density in the W14 catalog, and the right panels show this work.  The \citetalias{williams2014} catalog shows clear striping at the locations of the ACS chip gap, which was due to their exclusion of the full stack of 6-band imaging in these regions when performing the photometry.  

\begin{figure}
    \centering
    \includegraphics[width=3.1in]{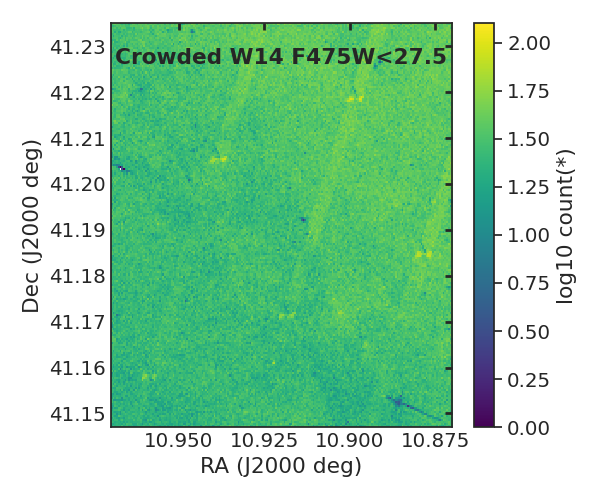}
    \includegraphics[width=3.1in]{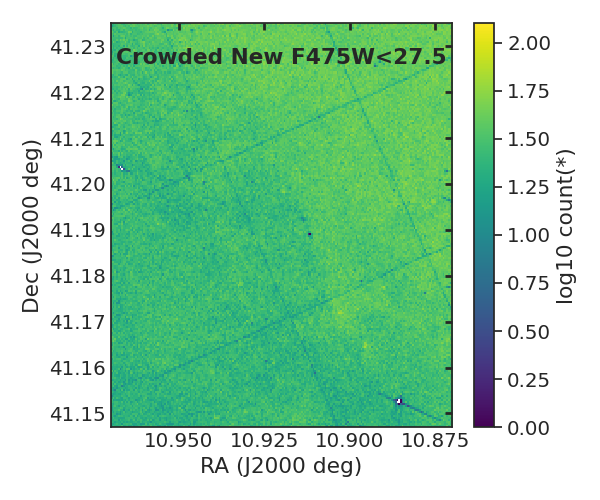}
    \includegraphics[width=3.1in]{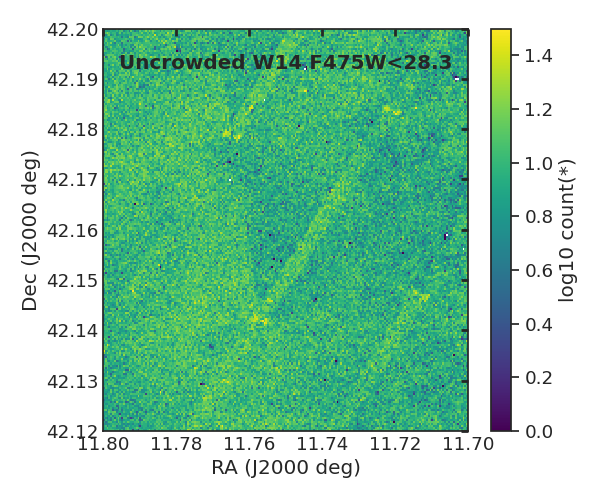}
    \includegraphics[width=3.1in]{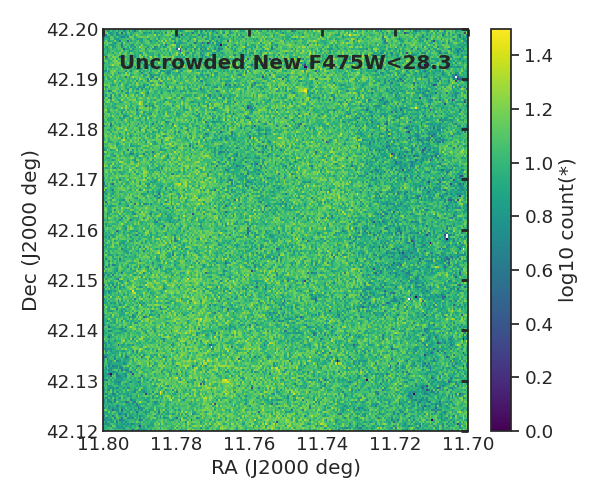}
    \caption{Same as Figure~\ref{fig:spatial_comparison}, but these maps are constructed only including stars with F475W<28.3, so that the patterns due to the improved depth in the new photometry are not visible.}
    \label{fig:matched_depths}
\end{figure}

This work still shows some spatial systematics, but they are mostly explained by the increased depth in areas where including overlapping exposures increased the number of detected stars.  In this portion of Brick 2, the map now has 12\% more stars, and this portion of Brick 22 map now has 40\% more stars due to the increase in exposure.  However, if we only include stars brighter than F475W of 28.3 when constructing the maps, as shown in Figure~\ref{fig:matched_depths}, most of the tiling structure vanishes from the new maps, other than the IR detector edge effects, whereas there is still striping visible in the maps from the old catalog.

Our new reduction shows improved depth in the areas of ACS overlap, especially in the ACS bands; however, the depth is also slightly
improved in the IR, likely due to the deeper ACS data helping with
deblending.  The new photometry also shows improved uniformity, in
that the ACS chip gaps are no longer as clearly visible, and most of the patterns
are easily attributable to the overlapping exposure pattern or
detector edge effects. % Meredith: permission to go into psf nonsense a tiny bit here? Ben: GRANTED! 

In the IR and optical, our new reduction shows the overlapping edges of
the IR detector.  Overlapping IR detector edges clearly impact the
spatial uniformity.  In regions where the photometry is not crowding
limited (e.g., Brick 22), the overlapping IR edges result in a slight
increase in depth (higher density of stars measured) in the IR.
However, in crowding-limited regions (e.g. Brick 2), this effect
appears to reverse, and the overlapping IR edges result in a slight
decrease in depth.  This decrease in depth in regions of overlapping
IR edges is visible in the optical in both the uncrowded (Brick 22)
and crowding limited (Brick 2) cases, suggesting that there are some photometry issues related to combining the edges of the IR detector. 

%One potential cause for this reduction of density of detections in overlapping regions could be a smaller number of spurious noise-related measurements where there are  more data.  The M31 background is bright enough that the sky noise is significant and can result in spurious detections of low signal to noise. This effect will be reduced when more exposures are available to lower the significance of the background noise level.  One clue that spurious low signal-to-noise detections are the cause is how the effect vanishes if only higher signal measurements are included.  In Fig. XXX, we show the same regionas Fig. XXX, but only including measurements of S/N>10. 

\begin{figure}
    \centering
    \includegraphics[width=6.1in]{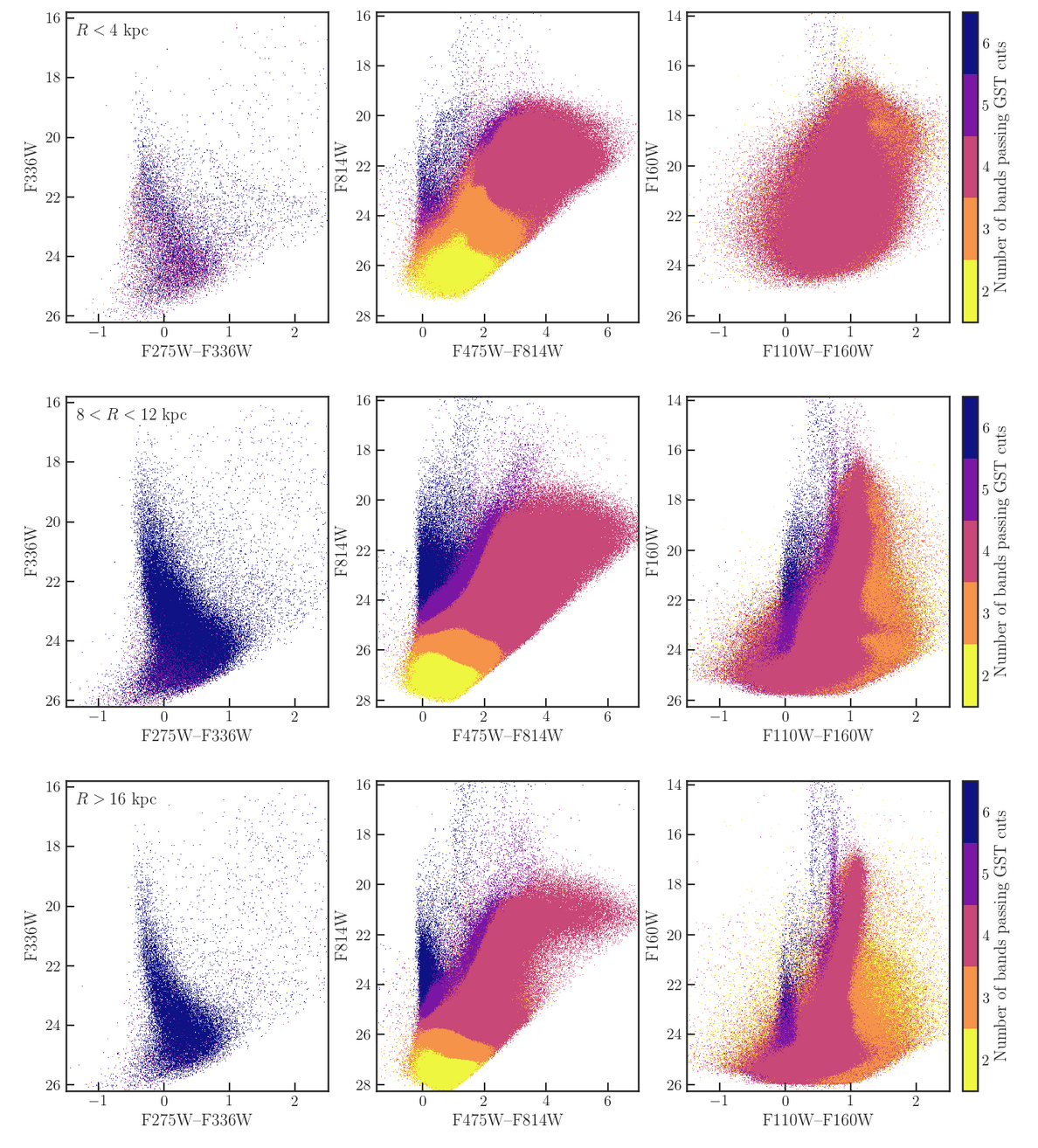}
    \caption{UV, optical, and IR plots showing the median number of bands with good measurements per star as a function of color and magnitude in 3 radial bins of the survey, representing high (top row), medium (middle row) and low (bottom row) stellar density regions. The CMD space is color-coded to show the number of bands with good measurements. In all panels, yellow, orange, magenta, purple, and blue points show stars with good measurements in two, three, four, five, and six bands, respectively. Left: UV CMD produced from all measurements where both bands passed our GST criteria. Center: optical CMD produced from all measurements where both bands passed our GST criteria. Right: IR CMD produced from all measurements where both bands passed our GST criteria.}
 \label{fig:filter_number}
\end{figure}

We also looked at the number of bands in which the stars typically had good measurements.  While the vast majority of the stars are well-measured in at least 4 bands, this number is a strong function of the type of star. In Figure~\ref{fig:filter_number}, we show CMDs of a small random set of stars color-coded by the number of bands in which they had high-quality (GST, for ``good star'') measurements.  Most red giant brach (RGB) stars are well-measured in the optical and IR, resulting in 4 well-measured bands.  Most UV-bright stars are also well-measured in the optical and IR, resulting in 6 well-measured bands.  Stars that are near the optical magnitude limit have 2 or 3 well-measured bands, and relatively blue RGB stars and relatively red main-sequence (MS) stars tend to have 5 well-measured bands.

We also note that the foreground contamination, seen as bright vertical plumes between F475W-F814W values of 1 and 3 and between F110W-F160W values of 0.5 and 0.8, is extremely similar to the W14 catalog.  This small foreground component is described in detail in Section 4 of W14. 

\subsection{Artificial Star Test Results}

\begin{figure}
    \centering
    \includegraphics[width=6.0in]{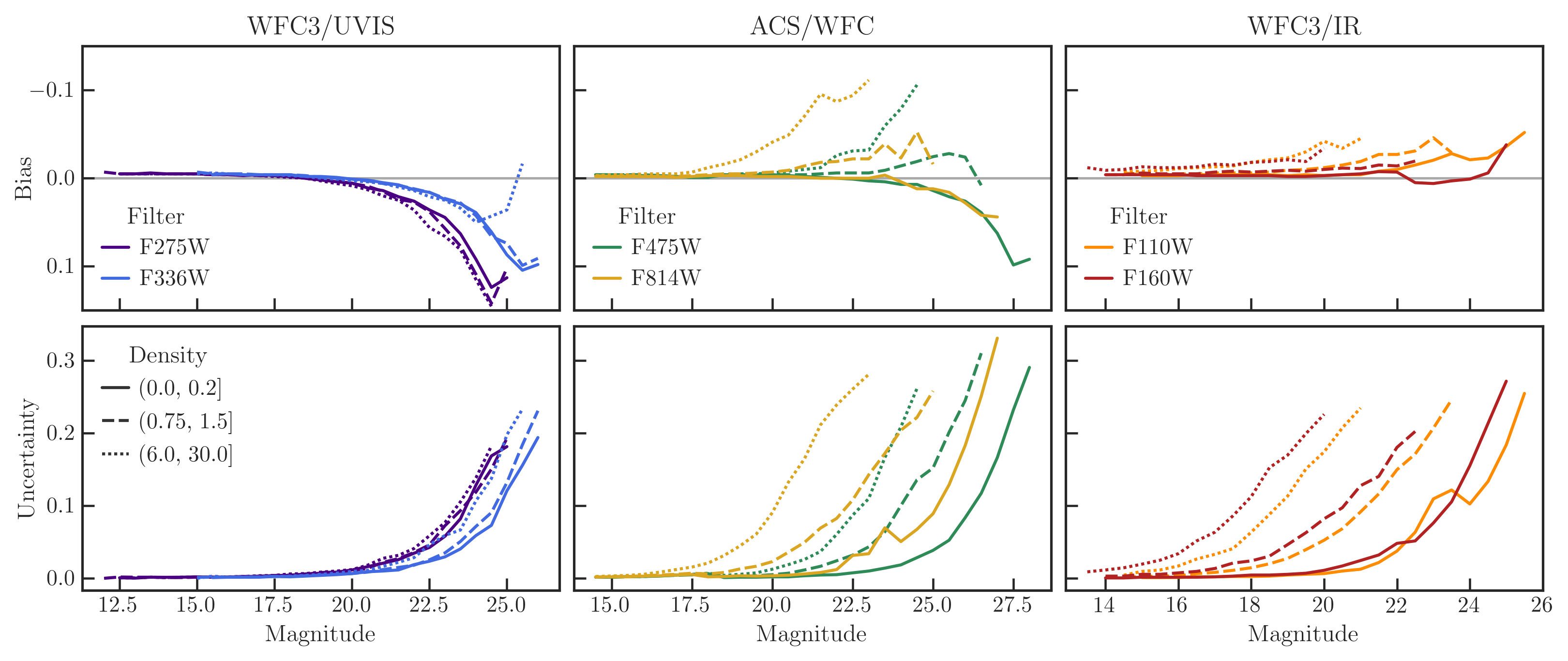}
    \caption{Bias and scatter results of our ASTs in all 6 bands as a function of input magnitude.   The bias is the median difference of the ASTs (output - input).  Scatter is calculated as the difference between the 84th and 16th percentile (output-input) differences divided by 2. In all bands, scatter increases with magnitude, and in the IR bands, the bias is for recovered stars to be brighter than input, as expected from crowding effects.  In uncrowded regions, the bias is for recovered stars to be fainter than input, suggesting slight background overestimation when optimizing for crowded fields.}
    \label{fig:ast_results}
\end{figure}

\begin{figure}
    \centering
    \includegraphics[width=6.0in]{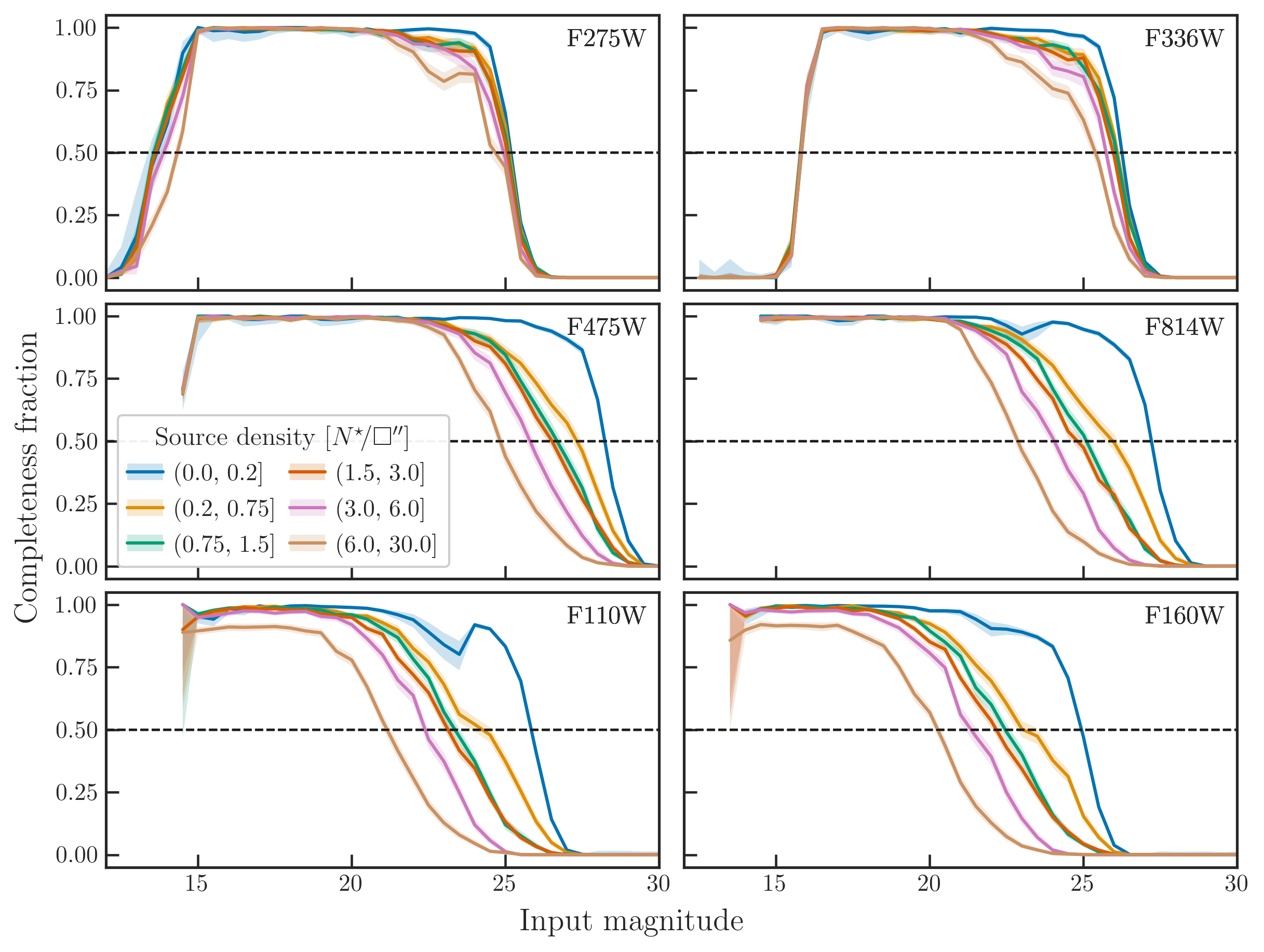}
    \caption{ Completeness as a function of magnitude for all 6 bands at six levels of stellar density.  Higher stellar densities show lower completeness due to crowding, with the effect showing up more strongly in redder bands. The statistics at the bright end are essentially unaffected by the small number of outliers detected.}
    \label{fig:completeness}
\end{figure}

Finally, the artificial star tests (ASTs) provide the most reliable measurement of the uncertainties, bias and completeness as a function of magnitude and stellar density in all 6 bands.  In Table~\ref{tab:ast_results}, we provide a simplified table of the results of our ASTs for easy use by readers.  However, the full AST output, including all of the output quality metrics, will be included in the data release on the MAST HLSP.   In Figure~\ref{fig:ast_results}, we show the residuals of the recovered ASTs as a function of input magnitude for all 6 bands.  In general, the results are well-behaved, showing increasing scatter and bias, and decreasing recoveries, at fainter magnitudes.  In Figure~\ref{fig:completeness}, we show the completeness fraction as a function of magnitude for all six bands at six levels of stellar density.  The redder bands are more affected by crowding than the bluer bands, as shown by the larger difference between the 50\% completeness magnitudes as the stellar density increased.  Because the population is dominated by red giant branch stars, and because HST has a broader point spread function in the red, crowding is a more significant in the images in the red bands.  We give the quantitative scatter, bias, and completeness for a range of magnitudes for each band and stellar density, in Tables~\ref{tab:ast_statistics} and \ref{tab:completeness}.  

The completeness reported in Table~\ref{tab:completeness} is for stars recovered using our GST criteria, which are the same as those defined in W14 (see their Table 3).  Thus it is limited mainly by signal-to-noise in the UV bands.  However, we note that because of the star forced PSF photometry in all bands at the locations of all stars, our catalog contains low S/N measurements in the UV of sources at much fainter fluxes.  For example, without requiring S/N$>$4 in F336W, our F336W 50\% completeness limit at low densities F336W$\sim$27.2, roughly a magnitude fainter than obtained when enforcing the GST S/N cut.  Furthermore, even when this cut is enforced, we see a gain of $\sim$0.1$-$0.2 mag in depth over W14 in the UV bands due to the inclusion of overlapping pointings.  Finally, in the optical and IR  we gain $\sim$0.2-0.4 mag and $\sim$0-0.7 mag in depth, respectively over W14 depending on band and stellar density, with the most gain at intermediate stellar densities.  The optical gain is primarily due to the inclusion of the overlapping pointings, while the IR gain is primarily due to improved deblending made possible by the improvement in the optical.

One interesting feature of the AST results is a small number of recovery anomalies for ASTs at bright magnitudes ($<$18) in the optical bands.  At those high fluxes, the stars are saturated in all but the very short "guard" exposures performed by the survey.  If such a bright star was not well-measured by that single exposure, there is a chance of poor photometry.  Our ASTs suggest that the probability of this kind of failure is 1.5\%.  Thus users should be aware that our catalog may contain some statistical outliers in the optical at such bright optical magnitudes.  Checking for rough consistency with Gaia for such bright stars is recommended for science that relies on such bright measurements.

\section{Discussion}

\begin{figure}
    \centering
    \includegraphics[width=6.0in]{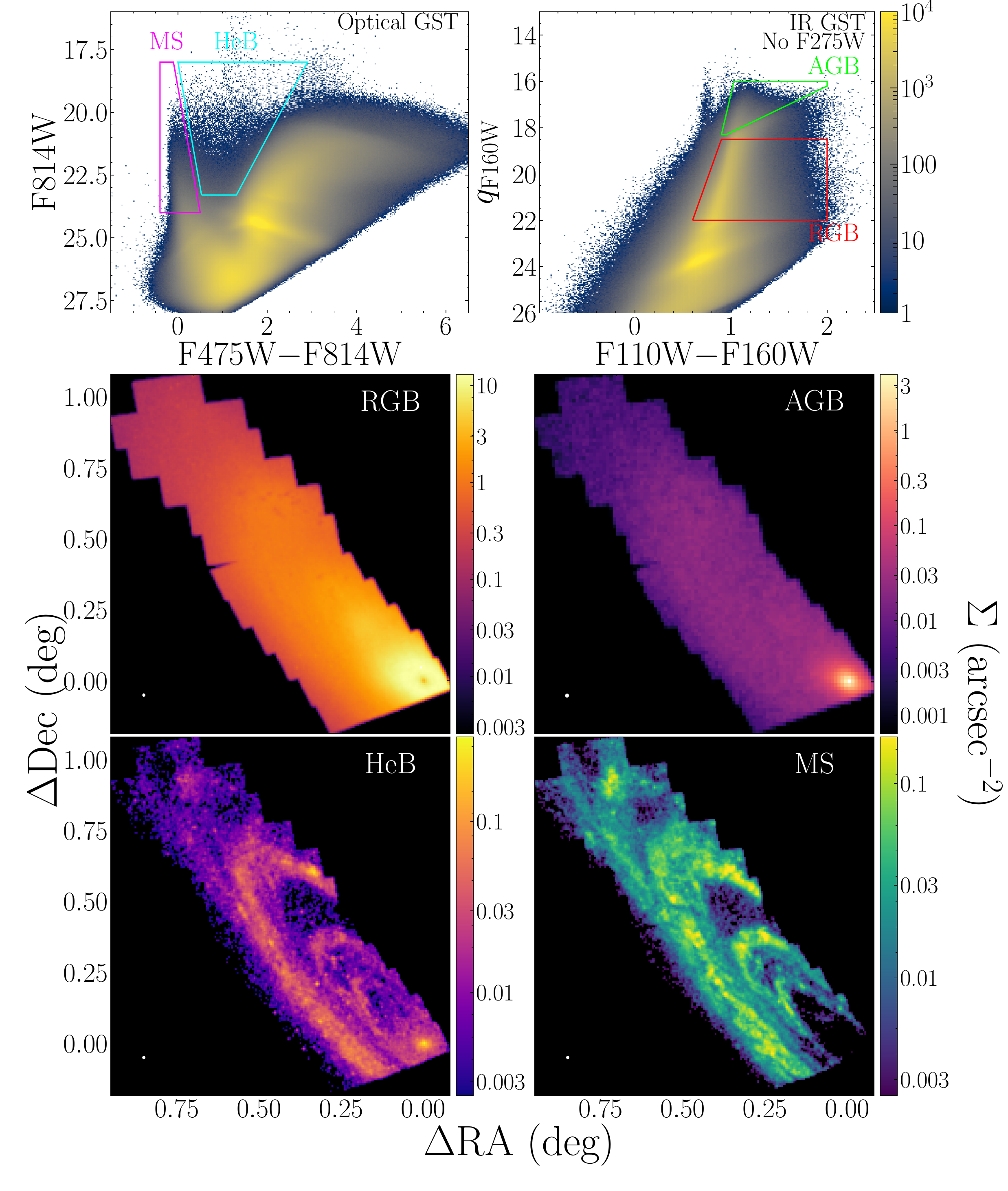}
    \caption{Maps of the spatial distribution of 4 CMD features that are proxies for different age populations. The top row provides a summary of the population selection used to create the maps.  The white dot in the lower left of each map shows the size of the smoothing kernel used.  The color bar is in units of stars per CMD bin. The left panel shows outlines of the areas of the optical CMD from which the MS (red polygon) and helium burning (cyan polygon) populations were extracted, while the right panel shows outlines of the areas of the infrared CMD from which the RGB (red polygon) and AGB (green polygon) populations were extracted.  The MS probes $\sim$3-200 Myr, and the HeB is probes $\sim$30-500 Myr.  The AGB probes $\sim$0.8-2 Gyr and the RGB probes $\gap$2 Gyr. The bottom 2 rows show the resulting maps showing that the 2 younger populations are more highly structured than the 2 older populations.}
    \label{fig:population_maps}
\end{figure}

Our new photometry has significantly lower spatial systematics as seen in the close-up spatial density maps described earlier.  This reduction in systematic spatial variations simplifies the production of stellar density maps for populations of interest.  We show 4 examples of such maps in Figure~\ref{fig:population_maps}.  These figures show the boundaries of 4 color-magnitude selection regions that isolate populations of 4 rough age ranges, following Smercina et al.\ (2023, ApJ, submitted).  The top row of this figure provides a summary of the population selection used to create the maps.  The left panel shows outlines of the areas of the optical CMD from which the MS (red polygon) and helium burning (cyan polygon) populations were extracted, while the middle panel shows outlines of the areas of the infrared CMD from which the RGB (red polygon) and asymptotic giant branch (AGB; cyan polygon) populations were extracted.  The  distribution of ages of stars in these regions as determined from a model population of constant star formation rate are as follows. The MS selection probes $\sim$3-200 Myr old population, and the HeB selection probes $\sim$30-500 Myr old populations.  The upper-right shows the infrared selection of the older populations, where the AGB probes $\sim$0.8-2 Gyr and the RGB probes $\gap$2 Gyr.

These selection panels are followed by the spatial maps of the stars that fall in those selection regions.  The virtually seamless maps provide further confirmation of the high fidelity of this legacy photometry.

These maps also provide some insight into the structural evolution of the M31 disk.  While the detailed star formation history of the disk has been measured with the PHAT data \citep{lewis2015,williams2017}, these maps provide a lower time resolution look that confirms the overall results of those detailed measurements.  For example, one of the major results from \citet{lewis2015} was that the ring features appear to be long-lived, appearing clearly for all ages back to 600 Myr in age.  We see in these maps that indeed the populations that represent ages younger than about 1 Gyr, the HeB and MS features, are largely confined to the ring features.  Furthermore, \citet{williams2015,williams2017} found that there appeared to be a disk-wide burst of star formation 2-4 Gyr ago, a feature that was also seen across previous sparse HST pointings \citep{bernard2012,bernard2015}.  These maps show that the populations older than a Gyr, AGB and RGB, are widespread, showing far less structure, consistent with such a widespread event near that age boundary.

%XXX Do we want to put the AGB/RGB in this paper?
%To further investigate the evolution of the M31 disk using population mapping, we generated a map of the AGB/RGB ratio in Figure~\ref{agb2rgb}.  If we assume the RGB stars are all very old ($\gap$10 Gyr) and well-mixed, this ratio is sensitive to mixing that has occured on the timescale of the ages of the AGB stars, essentially normalizing it by the overall disk structure, defined by the RGB.  This map shows ...

\section{Conclusions}

We have updated our technique for measuring photometry from the HST PHAT survey imaging of the northern half of the M31 disk.  This legacy photometry catalog was produced using all of the overlapping images at every location in the survey in order to optimize depth and minimize systematics caused by independent measurements of each pointing, which showed clear spatial systematics when combined into a single catalog.

Comparisons between this updated photometry and the previous release \citepalias{williams2014} show improvement in both depth and spatial consistency in the optical and near infrared, along with updated calibrations and CTE corrections in the near ultraviolet.  Furthermore, maps of the populations from this updated photometry catalog are virtually seamless and show clear structural evolution, in particular between populations associated with ages $\gap$1 Gyr ago and $\lap$1 Gyr ago, consistent with the possibility of a significant event in the history of M31 a few Gyr ago, and long lived star forming structures for about the past Gyr.

Support for this work was provided by NASA through grant GO-12055 from the Space Telescope Science Institute, which is operated by AURA, Inc., under NASA contract NAS 5-26555.

%\clearpage

\bibliographystyle{aasjournal}
\bibliography{apjmnemonic,references}
\clearpage

\begin{sidewaystable}
    \centering
    \caption{Simplified table of our photometry for easy use.  Give are the positions, Vega magnnitudes,   The first 10 lines are shown here.  The full 138 million line version is available in electronic formal in the supplemental data.}
    \begin{tabular}{cccccccccccccccccccc}
    \hline\hline
RA (J2000) & DEC (J2000) & F275W & SNR & G & F336W & SNR & G & F475W & SNR & G & F814W & SNR & G & F110W & SNR & G & F160W & SNR & G\\ % inserts table %heading
\hline
10.57619538 & 41.24338318 & 99.999 & -0.2 & 0 & 26.952 & 2.1 & 0 & 24.988 & 23.3 & 1 & 23.254 & 28.5 & 1 & 99.999 & 0.0 & 0 & 99.999 & 0.0 & 0 \\
10.57620236 & 41.24336347 & 29.530 & 0.1 & 0 & 27.954 & 0.9 & 0 & 25.190 & 18.7 & 1 & 23.422 & 24.5 & 1 & 99.999 & 0.0 & 0 & 99.999 & 0.0 & 0 \\
10.57623847 & 41.24342833 & 26.121 & 1.5 & 0 & 99.999 & -0.4 & 0 & 26.978 & 4.5 & 1 & 24.184 & 12.9 & 1 & 99.999 & 0.0 & 0 & 99.999 & 0.0 & 0 \\
10.57623862 & 41.24345726 & 99.999 & -0.4 & 0 & 99.999 & -2.4 & 0 & 26.256 & 11.0 & 1 & 24.611 & 11.5 & 1 & 99.999 & 0.0 & 0 & 99.999 & 0.0 & 0 \\
10.57627747 & 41.24343760 & 99.999 & -0.4 & 0 & 99.999 & -0.7 & 0 & 25.951 & 13.4 & 1 & 24.117 & 15.5 & 1 & 99.999 & 0.0 & 0 & 99.999 & 0.0 & 0 \\
10.57630315 & 41.24360886 & 26.079 & 1.3 & 0 & 99.999 & -0.8 & 0 & 26.549 & 7.9 & 1 & 24.276 & 11.5 & 1 & 99.999 & 0.0 & 0 & 99.999 & 0.0 & 0 \\
10.57631263 & 41.24340316 & 99.999 & -1.5 & 0 & 29.410 & 0.2 & 0 & 26.995 & 5.6 & 0 & 25.168 & 6.8 & 0 & 99.999 & 0.0 & 0 & 99.999 & 0.0 & 0 \\
10.57631284 & 41.24333622 & 99.999 & -1.0 & 0 & 27.143 & 1.6 & 0 & 26.957 & 6.1 & 1 & 24.279 & 10.9 & 1 & 99.999 & 0.0 & 0 & 99.999 & 0.0 & 0 \\
10.57631873 & 41.24347745 & 99.999 & -0.8 & 0 & 99.999 & -1.1 & 0 & 25.653 & 19.0 & 1 & 22.850 & 47.2 & 1 & 99.999 & 0.0 & 0 & 99.999 & 0.0 & 0 \\
10.57631976 & 41.24358853 & 26.863 & 1.0 & 0 & 99.999 & -0.0 & 0 & 26.260 & 8.2 & 1 & 24.131 & 13.2 & 1 & 99.999 & 0.0 & 0 & 99.999 & 0.0 & 0 \\
    \end{tabular}
    \label{tab:photometry_table}
\end{sidewaystable}

\begin{sidewaystable}
    \centering
    \tiny
    \caption{Simplified table of artificial star test results for easy use.  The first 10 lines are shown here.  The full 250,000 line version is available in electronic format in the supplemental data.}
    \begin{tabular}{rrr|rrrl|rrrl|rrrl|rrrl|rrrl|rrrl}
    \hline\hline
RA\tablenotemark{\tiny{a}} & DEC\tablenotemark{\tiny{a}} & \multicolumn{1}{r|}{Density} & \multicolumn{4}{c|}{F275W} & \multicolumn{4}{c|}{F336W} & \multicolumn{4}{c|}{F475W} & \multicolumn{4}{c|}{F814W} & \multicolumn{4}{c|}{F110W} & \multicolumn{4}{c}{F160W} \\
~ & ~ & ~ & In &  Out-In &   SNR & G &  In &  Out-In &   SNR & G &  In &  Out-In &   SNR & G &  In &  Out-In &   SNR & G &  In &  Out-In &   SNR & G &  In &  Out-In &  SNR & G \\
\hline
10.576947 & 41.244183 &   22.60 &    43.584 &    99.999 &   0.0 &    0 &    40.252 &    99.999 &   0.0 &    0 &    38.015 &    99.999 &   0.0 &    0 &    34.591 &    99.999 &   0.0 &    0 &    33.414 &    99.999 &    0.0 &    0 &    32.665 &    99.999 &    0.0 &    0 \\
10.577140 & 41.243620 &   22.16 &    30.619 &    -2.680 &   0.3 &    0 &    32.122 &    -3.048 &   0.3 &    0 &    27.298 &    -1.064 &   6.2 &    0 &    21.233 &    -0.028 & 153.7 &    1 &    19.677 &     0.052 &  172.3 &    1 &    18.740 &     0.038 &  252.7 &    1 \\
10.577270 & 41.243775 &   13.24 &    14.998 &    -0.002 & 530.5 &    1 &    15.055 &    99.999 &   0.0 &    0 &    16.152 &     0.007 & 280.2 &    1 &    15.728 &    -0.001 & 300.5 &    1 &    15.743 &     0.003 & 2539.2 &    1 &    15.697 &     0.010 & 1920.5 &    1 \\
10.577306 & 41.243625 &    1.32 &    30.670 &    99.999 &   0.0 &    0 &    28.472 &    99.999 &   0.0 &    0 &    27.083 &    99.999 &   0.0 &    0 &    24.882 &    99.999 &   0.0 &    0 &    24.055 &    99.999 &    0.0 &    0 &    23.257 &    99.999 &    0.0 &    0 \\
10.577317 & 41.244550 &    5.44 &    15.589 &    -0.004 & 662.1 &    1 &    16.252 &    -0.008 & 862.7 &    1 &    18.038 &     0.008 & 115.4 &    1 &    18.498 &    -0.001 & 639.8 &    1 &    18.829 &     0.017 &  509.7 &    1 &    18.980 &     0.011 &  280.1 &    1 \\
10.577446 & 41.244082 &    3.32 &    13.455 &    99.999 &   0.0 &    0 &    14.108 &    99.999 &   0.0 &    0 &    15.882 &    -0.004 & 319.5 &    1 &    16.319 &     0.005 & 227.5 &    1 &    16.639 &     0.000 & 1671.3 &    1 &    16.781 &     0.003 & 1137.3 &    1 \\
10.577992 & 41.243590 &   14.92 &    32.914 &    99.999 &  -0.2 &    0 &    34.865 &    -8.325 &   3.0 &    0 &    31.415 &    -5.867 &  11.0 &    1 &    24.319 &    -0.947 &  14.6 &    1 &    20.972 &     0.096 &   49.0 &    1 &    19.135 &     0.251 &   91.2 &    1 \\
10.578305 & 41.244659 &   13.64 &    44.412 &    99.999 &   0.0 &    0 &    42.779 &    99.999 &   0.0 &    0 &    39.885 &    99.999 &   0.0 &    0 &    35.205 &    99.999 &   0.0 &    0 &    33.390 &    99.999 &    0.0 &    0 &    32.382 &    99.999 &    0.0 &    0 \\
10.578338 & 41.243374 &   11.84 &    41.622 &    99.999 &   0.0 &    0 &    41.681 &    99.999 &   0.0 &    0 &    39.023 &    99.999 &   0.0 &    0 &    33.136 &    99.999 &   0.0 &    0 &    30.001 &    99.999 &    0.0 &    0 &    28.528 &    99.999 &    0.0 &    0 \\
10.578630 & 41.243383 &   17.88 &    45.766 &    99.999 &   0.0 &    0 &    47.826 &    99.999 &   0.0 &    0 &    44.119 &    99.999 &   0.0 &    0 &    36.934 &    99.999 &   0.0 &    0 &    34.174 &    99.999 &    0.0 &    0 &    32.747 &    99.999 &    0.0 &    0 \\
\hline
\end{tabular}
    \label{tab:ast_results}
    \tablenotetext{\tiny{a}}{J2000}
\end{sidewaystable}

\begin{table}
    \centering
    \caption{Summary of artificial star statistics as a function of brightness and stellar density.}
\begin{tabular}{llrrrrr}
\hline\hline
    Density & Filter &  Magnitude &    Bias &  Uncertainty &  DOLPHOT &    Ratio \\
\hline
(0.0, 0.2] &  F275W &       12.5 & -0.0050 &       0.0007 &    0.001 & 0.6800 \\
(0.0, 0.2] &  F275W &       13.0 & -0.0050 &       0.0003 &    0.001 & 0.2600 \\
(0.0, 0.2] &  F275W &       13.5 & -0.0060 &       0.0015 &    0.001 & 1.5000 \\
(0.0, 0.2] &  F275W &       14.0 & -0.0050 &       0.0014 &    0.001 & 1.3800 \\
(0.0, 0.2] &  F275W &       14.5 & -0.0050 &       0.0015 &    0.001 & 1.5000 \\
(0.0, 0.2] &  F275W &       15.0 & -0.0050 &       0.0020 &    0.001 & 2.0000 \\
(0.0, 0.2] &  F275W &       15.5 & -0.0040 &       0.0015 &    0.001 & 1.5000 \\
(0.0, 0.2] &  F275W &       16.0 & -0.0040 &       0.0017 &    0.002 & 0.8700 \\
(0.0, 0.2] &  F275W &       16.5 & -0.0030 &       0.0025 &    0.002 & 1.2500 \\
(0.0, 0.2] &  F275W &       17.0 & -0.0040 &       0.0025 &    0.003 & 0.8333 \\
(0.0, 0.2] &  F275W &       17.5 & -0.0030 &       0.0039 &    0.004 & 0.9750 \\
(0.0, 0.2] &  F275W &       18.0 & -0.0020 &       0.0040 &    0.004 & 1.0000 \\
(0.0, 0.2] &  F275W &       18.5 &  0.0000 &       0.0060 &    0.006 & 1.0000 \\
(0.0, 0.2] &  F275W &       19.0 &  0.0020 &       0.0070 &    0.007 & 1.0000 \\
(0.0, 0.2] &  F275W &       19.5 &  0.0040 &       0.0085 &    0.009 & 0.9444 \\
(0.0, 0.2] &  F275W &       20.0 &  0.0055 &       0.0115 &    0.011 & 1.0455 \\
(0.0, 0.2] &  F275W &       20.5 &  0.0110 &       0.0160 &    0.014 & 1.1429 \\
(0.0, 0.2] &  F275W &       21.0 &  0.0140 &       0.0208 &    0.019 & 1.0926 \\
(0.0, 0.2] &  F275W &       21.5 &  0.0210 &       0.0257 &    0.022 & 1.1682 \\
(0.0, 0.2] &  F275W &       22.0 &  0.0260 &       0.0345 &    0.032 & 1.0788 \\
(0.0, 0.2] &  F275W &       22.5 &  0.0360 &       0.0425 &    0.041 & 1.0366 \\
(0.0, 0.2] &  F275W &       23.0 &  0.0445 &       0.0579 &    0.056 & 1.0339 \\
(0.0, 0.2] &  F275W &       23.5 &  0.0630 &       0.0820 &    0.076 & 1.0789 \\
(0.0, 0.2] &  F275W &       24.0 &  0.0920 &       0.1284 &    0.110 & 1.1671 \\
(0.0, 0.2] &  F275W &       24.5 &  0.1240 &       0.1687 &    0.153 & 1.1026 \\
(0.0, 0.2] &  F275W &       25.0 &  0.1130 &       0.1815 &    0.200 & 0.9075 \\
(0.0, 0.2] &  F275W &       25.5 &  0.0480 &       0.2361 &    0.232 & 1.0177 \\
(0.0, 0.2] &  F275W &       26.0 & -0.2855 &       0.2675 &    0.245 & 1.0920 \\
(0.0, 0.2] &  F275W &       26.5 & -0.5140 &       0.0612 &    0.236 & 0.2593 \\
\hline
\end{tabular}
    \label{tab:ast_statistics}
\end{table}

\begin{table}
\centering
\caption{50\% completeness limits by stellar density (stars/square arcsec with 20$<$F814W$<$22).}
\label{tab:completeness}
\begin{tabular}{lrrrrrr}
\hline
\hline
Density &  F275W &  F336W &  F475W &  F814W &  F110W &  F160W \\
\hline
(0, 0.2]     &  25.18 &  26.25 &  28.24 &  27.21 &  25.84 &  24.94 \\
(0.2, 0.75]  &  25.13 &  26.10 &  27.32 &  25.96 &  24.25 &  23.04 \\
(0.75, 1.5]  &  25.09 &  26.07 &  26.68 &  25.10 &  23.36 &  22.46 \\
(1.5, 3.0]   &  25.05 &  25.97 &  26.52 &  24.80 &  23.13 &  22.16 \\
(3.0, 6.0]   &  24.94 &  25.74 &  25.80 &  24.05 &  22.39 &  21.34 \\
(6.0, 30.0]  &  24.67 &  25.39 &  24.83 &  22.86 &  21.15 &  20.26 \\
\hline
\end{tabular}
\end{table}

\clearpage

%\begin{figure}
%%    \includegraphics[width=6.0in]{f475w_f814w_gst_cmd.png}
%    \caption{Example optical color-magnitude diagram from our new catalog, showing the same %feaures seen in W14.   The color bar is in units of the log of the number of stars per CMD bin. The blue plume is the upper-main-sequence.  The wide red plume is the red giant branch.  The elongated peak near 2, 24 is the red clump, and above that, near 2, 23 is the AGB bump.}
%    \label{fig:example_cmd}
%\end{figure}

% MJD: this is the caption for this figure from W14:
% "Map of the six levels of model stellar densities used to define the six regions that were used to make the different CMDs shown in Figures 11–18. Numbers refer to the log of the stellar density of red giant branch stars. The numbers are relative and determined from a model disk, making the normalization irrelevant. The areas covered by the defined regions are 770, 481 349, 113, 106, and 31 arcmin2 from lowest to highest stellar density, respectively. White squares mark the centers of the fields where artificial star tests were performed to probe a range of stellar densities. We divided our results in this way to account for the changing completeness limits with stellar density."

\clearpage

\clearpage

\clearpage

\end{document}